\documentclass[ALICE,manyauthors]{cernphprep}
\usepackage[comma,square,numbers,sort&compress]{natbib}
\usepackage{hyperref}
\usepackage{lineno}
\usepackage{xcolor}
\usepackage{euscript}
\usepackage[scientific-notation=true]{siunitx}
\usepackage[greek,english]{babel}
\usepackage{multirow}
\usepackage[Symbolsmallscale]{upgreek}
\usepackage{multirow}

\usepackage[T1]{fontenc}
\usepackage{orcidlink}

\begin{document}
%

\newcommand{\pp}           {pp\xspace}
\newcommand{\ppbar}        {\mbox{$\mathrm {p--\overline{p}}$}\xspace}
\newcommand{\XeXe}         {\mbox{Xe--Xe}\xspace}
\newcommand{\PbPb}         {\mbox{Pb--Pb}\xspace}
\newcommand{\pAa}           {\mbox{pA}\xspace}
\newcommand{\pPb}          {\mbox{p--Pb}\xspace}
\newcommand{\AuAu}         {\mbox{Au--Au}\xspace}
\newcommand{\dAu}          {\mbox{d--Au}\xspace}
\newcommand{\pP}{\ensuremath{\mbox{p--p}}\,}


\newcommand{\kstar}        {\ensuremath{k^*}\xspace}
\newcommand{\rstar}     {\ensuremath{r^{*}}\xspace}

\newcommand{\snn}          {\ensuremath{\sqrt{s_{\mathrm{NN}}}}\xspace}
\newcommand{\pt}           {\ensuremath{p_{\rm T}}\xspace}
\newcommand{\meanpt}       {$\langle p_{\mathrm{T}}\rangle$\xspace}
\newcommand{\ycms}         {\ensuremath{y_{\rm CMS}}\xspace}
\newcommand{\ylab}         {\ensuremath{y_{\rm lab}}\xspace}
\newcommand{\etarange}[1]  {\mbox{$\left | \eta \right |~<~#1$}}
\newcommand{\yrange}[1]    {\mbox{$\left | y \right |~<~#1$}}
\newcommand{\dndy}         {\ensuremath{\mathrm{d}N_\mathrm{ch}/\mathrm{d}y}\xspace}
\newcommand{\dndeta}       {\ensuremath{\mathrm{d}N_\mathrm{ch}/\mathrm{d}\eta}\xspace}
\newcommand{\avdndeta}     {\ensuremath{\langle\dndeta\rangle}\xspace}
\newcommand{\dNdy}         {\ensuremath{\mathrm{d}N_\mathrm{ch}/\mathrm{d}y}\xspace}
\newcommand{\Npart}        {\ensuremath{N_\mathrm{part}}\xspace}
\newcommand{\Ncoll}        {\ensuremath{N_\mathrm{coll}}\xspace}
\newcommand{\dEdx}         {\ensuremath{\textrm{d}E/\textrm{d}x}\xspace}
\newcommand{\RpPb}         {\ensuremath{R_{\rm pPb}}\xspace}

\newcommand{\nineH}        {$\sqrt{s}~=~0.9$~Te\kern-.1emV\xspace}
\newcommand{\seven}        {$\sqrt{s}~=~7$~Te\kern-.1emV\xspace}
\newcommand{\onethree}        {$\sqrt{s}~=~13$~Te\kern-.1emV\xspace}
\newcommand{\twoH}         {$\sqrt{s}~=~0.2$~Te\kern-.1emV\xspace}
\newcommand{\twosevensix}  {$\sqrt{s}~=~2.76$~Te\kern-.1emV\xspace}
\newcommand{\five}         {$\sqrt{s}~=~5.02$~Te\kern-.1emV\xspace}
\newcommand{\twosevensixnn}{$\sqrt{s_{\mathrm{NN}}}~=~2.76$~Te\kern-.1emV\xspace}
\newcommand{\fivenn}       {$\sqrt{s_{\mathrm{NN}}}~=~5.02$~Te\kern-.1emV\xspace}
\newcommand{\LT}           {L{\'e}vy-Tsallis\xspace}
\newcommand{\GeVc}         {Ge\kern-.1emV$/c$\xspace}
\newcommand{\gevc}         {Ge\kern-.1emV$/c$\xspace}
\newcommand{\MeVc}         {Me\kern-.1emV$/c$\xspace}
\newcommand{\mevc}         {Me\kern-.1emV$/c$\xspace}
\newcommand{\GeVmass}      {Ge\kern-.1emV$/c^2$\xspace}
\newcommand{\MeVmass}      {Me\kern-.1emV$/c^2$\xspace}
\newcommand{\mevcc}      {Me\kern-.1emV$/c^2$\xspace}
\newcommand{\gevcc}      {Ge\kern-.1emV$/c^2$\xspace}
\newcommand{\lumi}         {\ensuremath{\mathcal{L}}\xspace}

\newcommand{\ITS}          {\rm{ITS}\xspace}
\newcommand{\TOF}          {\rm{TOF}\xspace}
\newcommand{\ZDC}          {\rm{ZDC}\xspace}
\newcommand{\ZDCs}         {\rm{ZDCs}\xspace}
\newcommand{\ZNA}          {\rm{ZNA}\xspace}
\newcommand{\ZNC}          {\rm{ZNC}\xspace}
\newcommand{\SPD}          {\rm{SPD}\xspace}
\newcommand{\SDD}          {\rm{SDD}\xspace}
\newcommand{\SSD}          {\rm{SSD}\xspace}
\newcommand{\TPC}          {\rm{TPC}\xspace}
\newcommand{\TRD}          {\rm{TRD}\xspace}
\newcommand{\VZERO}        {\rm{V0}\xspace}
\newcommand{\VZEROA}       {\rm{V0A}\xspace}
\newcommand{\VZEROC}       {\rm{V0C}\xspace}
\newcommand{\Vdecay} 	   {\ensuremath{V^{0}}\xspace}

\newcommand{\ee}           {\ensuremath{e^{+}e^{-}}} 
\newcommand{\pip}          {\ensuremath{\pi^{+}}\xspace}
\newcommand{\pim}          {\ensuremath{\pi^{-}}\xspace}
\newcommand{\kap}          {\ensuremath{\rm{K}^{+}}\xspace}
\newcommand{\kam}          {\ensuremath{\rm{K}^{-}}}
\newcommand{\pbar}         {\ensuremath{\rm\overline{p}}\xspace}
\newcommand{\kzeros}        {\ensuremath{{\rm K}^{0}_{\rm{S}}}\xspace}
\newcommand{\kzerobar}     {\ensuremath{\rm \overline{K}^0}}
\newcommand{\lmb}          {\ensuremath{\Lambda}\xspace}
\newcommand{\almb}         {\ensuremath{\overline{\Lambda}}\xspace}
\newcommand{\prot}         {\ensuremath{\rm{p}}\xspace}
\newcommand{\aprot}         {\ensuremath{\rm{\overline{p}}}\xspace}
\newcommand{\n}         {\ensuremath{\rm{n}}\xspace}
\newcommand{\an}         {\ensuremath{\rm{\overline{n}}}\xspace}
\newcommand{\kbar}         {\ensuremath{\rm\overline{K}}\xspace}

\newcommand{\Om}           {\ensuremath{\Omega^-}\xspace}
\newcommand{\Mo}           {\ensuremath{\overline{\Omega}^+}\xspace}
\newcommand{\X}            {\ensuremath{\Xi^-}\xspace}
\newcommand{\Ix}           {\ensuremath{\overline{\Xi}^+}\xspace}
\newcommand{\Xis}          {\ensuremath{\Xi^{\pm}}\xspace}
\newcommand{\Oms}          {\ensuremath{\Omega^{\pm}}\xspace}
\newcommand{\SigZ}            {\ensuremath{\Sigma^0}\xspace}
\newcommand{\aSigZ}            {\ensuremath{\overline{\Sigma^0}}\xspace}
\newcommand{\antik}   {$\mathrm{\overline{K}}\,$}

\newcommand{\Ledn}         {Lednick\'y--Lyuboshits\xspace}
\newcommand{\chiEFT}       {\ensuremath{\chi}\rm{EFT}\xspace}
\newcommand{\ks}     {\ensuremath{k^{*}}\xspace}
\newcommand{\rs}     {\ensuremath{r^{*}}\xspace}
\newcommand{\mt}     {\ensuremath{m_{\mathrm{T}}}\xspace}
\newcommand{\Cth}           {C_\mathrm{th}\xspace}
\newcommand{\Cexp}           {C_\mathrm{exp}\xspace}
\newcommand{\CF}           {\ensuremath{C(\ks)}\xspace}
\newcommand{\Cgen}           {\ensuremath{C_{\mathrm{gen}}(\ks)}\xspace}

\newcommand{\Sr}            {\ensuremath{S(\rs)}\xspace}
\newcommand{\BBar}            {\ensuremath{\rm{B}\mbox{--}\rm{\overline{B}}}\xspace}
\newcommand{\SPi}         {\ensuremath{\uppi\Sigma}\xspace}
\newcommand{\kbarN}         {\ensuremath{\rm\overline{K}N}\xspace}
\newcommand{\LL}            {\ensuremath{\lmb\mbox{--}\lmb}\xspace}
\newcommand{\pprot}            {\ensuremath{\prot\mbox{--}\prot}\xspace}
\newcommand{\pL}            {\ensuremath{\prot\mbox{--}\lmb}\xspace}

\newcommand{\LK}{$\Lambda \mbox{--}{\mathrm{K}}$\xspace} 
\newcommand{\LAK}{$\Lambda \mbox{--}{\mathrm{\overline{K}}}$\xspace} 
\newcommand{\LKzero}{$\Lambda \mbox{--}{\mathrm{K^0 _S}}$\xspace} 

\newcommand{\AKN}{${\mathrm{\overline{K}}}$N\xspace} 
\newcommand{\aKN}{${\mathrm{\overline{K}}}$N\xspace} 
\newcommand{\KbarN}{${\mathrm{\overline{K}}}$N\xspace} 
\newcommand{\Kminusp}{\ensuremath{\mathrm{K}^-}p\xspace}
\newcommand{\Kminusd}{\ensuremath{\mathrm{K}^-}d\xspace}

\newcommand{\XiK}{\ensuremath{\Xi \mbox{--}\mathrm{K}}\xspace}
\newcommand{\XiPi}{\ensuremath{\Xi \mbox{--}\uppi}\xspace}
\newcommand{\Xipi}{\ensuremath{\Xi \mbox{--}\uppi}\xspace}

\newcommand{\LKInt}{$\Lambda{\mathrm{K}}$\xspace} 
\newcommand{\LAKInt}{$\Lambda{\mathrm{\overline{K}}}$\xspace}
\newcommand{\LKMinInt}{$\Lambda$K$^-$\,}
\newcommand{\LKPlusInt}{$\Lambda$K$^+$\,}

\newcommand{\LKPlus}{$\Lambda \mbox{--}$K$^+$\,}
\newcommand{\ALKMin}{$\overline{\Lambda} \mbox{--}$K$^-$\,}
\newcommand{\LKMin}{$\Lambda \mbox{--}$K$^-$\,}
\newcommand{\ALKPlus}{$\overline{\Lambda} \mbox{--}$K$^+$\,}
\newcommand{\LKpair}{$\Lambda \mbox{--} $K$^+\oplus \overline{\Lambda} \mbox{--} $K$^-$\xspace}
\newcommand{\ALKpair}{$\Lambda \mbox{--} $K$^-\oplus \overline{\Lambda} \mbox{--} $K$^+$\xspace}
\newcommand{\KpMin}{K$^-p$\xspace}

\newcommand{\SAK}{$\Sigma\mathrm{\overline{K}}$\xspace} 
\newcommand{\SK}{$\Sigma \mathrm{K}$\xspace}

\newcommand{\Imscatt}            {\ensuremath{\Im f_0}\xspace}
\newcommand{\Ifzero}            {\ensuremath{\Im f_0}\xspace}
\newcommand{\Rescatt}            {\ensuremath{\Re f_0}\xspace}
\newcommand{\Rfzero}            {\ensuremath{\Re f_0}\xspace}
\newcommand{\effran}            {\ensuremath{d_0}\xspace}

\newcommand{\ImscattAmpl}            {\ensuremath{\Im f}\xspace}
\newcommand{\RescattAmpl}            {\ensuremath{\Re f}\xspace}

\newcommand{\XRes}          {\ensuremath{\Xi\mathrm{(1620)}}\xspace}
\newcommand{\XResNovanta}          {\ensuremath{\Xi\mathrm{(1690)}}\xspace}
\newcommand{\XResVenti}          {\ensuremath{\Xi\mathrm{(1820)}}\xspace}
\newcommand{\XResTrenta}          {\ensuremath{\Xi\mathrm{(1530)}}\xspace}
\newcommand{\XResDuemila}          {\ensuremath{\Xi\mathrm{(2500)}}\xspace}

\newcommand{\Xic}          {\ensuremath{\Xi_{c}}\xspace}

\newcommand{\XicZero}          {\ensuremath{\Xi_{c} ^0}\xspace}

\newcommand{\inlineannotation}[1]{%
    \hspace{2em}\parbox{0.9\linewidth}{\footnotesize\textit{#1}}%
}

\newcommand{\XiMinPi}{\ensuremath{\Xi^-  \mbox{--}\uppi^+}\xspace}
\newcommand{\XiMinK}{\ensuremath{\Xi^-  \mbox{--}\mbox{K$^{+}$}}\xspace}
\newcommand{\XiPPi}{\ensuremath{\overline{\Xi}^+  \mbox{--}\uppi^-}\xspace}
\newcommand{\XiPK}{\ensuremath{\overline{\Xi}^+  \mbox{--}\mbox{K$^{-}$}}\xspace}


\newcommand{\rUNOXipi}{$r_\mathrm{1} = 1.19\pm 0.04  \text{ fm}$}
\newcommand{\rDOSXipi}{$r_\mathrm{2} = 3.16\pm 0.04  \text{ fm}$}
\newcommand{\alphaSXipi}{$\lambda_\mathrm{S} = 0.97$}
\newcommand{\omegaSXipi}{$\omega_\mathrm{S} = 0.74$}

\newcommand{\rUNOXiK}{$r_\mathrm{1} = 1.00\pm 0.04  \text{ fm}$}
\newcommand{\rDOSXiK}{$r_\mathrm{2} = 2.45\pm 0.04  \text{ fm}$}
\newcommand{\alphaSXiK}{$\lambda_\mathrm{S} = 0.97$}
\newcommand{\omegaSXiK}{$\omega_\mathrm{S} = 0.86$}

\newcommand{\rcoremTXipi}{
$r_\mathrm{core}(\left<m_\mathrm{T}\right> =1.55\,\,\mathrm{Ge\kern-.2emV/}c^2)=0.94\pm0.04\text{ fm}$
}
\newcommand{\rcoremTXiK}{
$r_\mathrm{core}(\left<m_\mathrm{T}\right> =1.69\,\,\mathrm{Ge\kern-.2emV/}c^2)=0.89\pm0.04\text{ fm}$
}

\newcommand{\RfzeroXiK}{$\Rfzero = -0.61 \pm 0.02 (\mathrm{stat.}) \pm 0.07 (\mathrm{syst.}) \text{ fm}~$}
\newcommand{\IfzeroXiK}{$\Ifzero = 0.41 \pm 0.04 (\mathrm{stat.}) \pm 0.01 (\mathrm{syst.}) \text{ fm}~$}

\newcommand{\Sminusone}            {S$~= -1~$} 
\newcommand{\Sminustwo}            {S$~= -2~$} 

\begin{titlepage}
\PHyear{2025}       
\PHnumber{215}      
\PHdate{17 Sep}  
\title{Study of the interaction between $\Xi$ baryons and light mesons via femtoscopy at the LHC}
\ShortTitle{Femtoscopy on $\Xi$--K,  $\Xi$--$\uppi$ pairs}   

\Collaboration{ALICE Collaboration\thanks{See Appendix~\ref{app:collab} for the list of collaboration members}}
\ShortAuthor{ALICE Collaboration} 

\begin{abstract}
Meson-baryon systems with strangeness content provide a unique laboratory for investigating the strong interaction and testing theoretical models of hadron structure and dynamics.
In this work, the measured correlation functions for 
oppositely charged \XiK and \XiPi pairs
obtained in high-multiplicity \pp collisions at \onethree at the LHC are presented.
For the first time, high-precision data on the \XiK interaction are delivered at small relative momenta.
The scattering lengths, extracted via the \Ledn expression of the pair wavefunction, indicate a repulsive and a shallow attractive strong interaction for the \XiK and \XiPi systems, respectively.
The \XRes and \XResNovanta states are observed in the \XiPi correlation function and their properties, mass and width, are determined.
These measurements are 
in agreement with other available results.
Such high-precision data can help refine the understanding of these resonant states, provide stronger constraints for chirally motivated potentials, and address the key challenge of describing the coupled-channel dynamics that may give rise to molecular configurations.
\end{abstract}
\end{titlepage}

\setcounter{page}{2} 

\section{Introduction}

Understanding the strong force acting at the hadronic level is fundamental for the development of Quantum Chromodynamics (QCD) in the non-perturbative regime.
The experimental and theoretical efforts in the study of the hadron spectrum over several decades revealed
features of QCD which lead to the appearance of exotic states
beyond the conventional classification of hadrons with
two and three valence quarks (mesons and baryons). 
This scenario 
is particularly evident for the charm sector, with several observations of tetra- and penta-quark states at high-energy \pp and e$^+$e$^-$ collider experiments~\cite{Belle:X3872,LHCb:Penta}.
In the light sector, where u, d and s quarks are considered, several predictions for molecular states exist, in particular for the meson-baryon sectors with strangeness \Sminusone~\cite{Mai:L1405Review,Jido:2003cbL1405,Magas:2005vuL1405,Oller:2000fjL1405,Zychor:2007gfL1405Exp,PhysRevC.87.025201L1405Exp,PhysRevLett.112.082004L1405Exp} and \Sminustwo~\cite{ALK_Ramos,Feijoo:MagasXiPi,BelleXi,LambdaK_Vale,LambdaK_Vale2}. The origin of these hadronic molecules lies in the characteristic coupled-channel dynamics occurring among meson-baryon pairs sharing the same quantum numbers. 
This coupled-channel interplay 
enables at the level of the interaction possible transitions, occurring both on- and off-shell, 
from one channel to the other. 
An example of a molecular state, the only one widely accepted so far, is the $\Lambda(1405)$, whose double-pole structure has been assessed to be arising from the $\uppi\Sigma-\kbarN$ coupling~\cite{Mai:L1405Review}. 
A stranger counterpart of the $\Lambda(1405)$ is represented by the \XRes in the \Sminustwo meson-baryon sector, whose structure is still rather unclear. This state was observed for the first time in 2019 by the Belle collaboration in the \XiPi channel~\cite{BelleXi}.
Along with the \XRes, also the \XResNovanta can be interpreted as a molecular state due to its proximity to the $\Sigma\kbar$ channels~\cite{LambdaK_Vale,Feijoo:MagasXiPi}. 
Thanks to a larger amount of spectroscopy data, in both \LKMin~\cite{LHCbXi16901820} and \XiPi~\cite{BelleXi} final states, the properties of this state, such as mass and width, are better known.

Effective field theories (EFTs), such as Chiral Perturbation Theory (ChPT)~\cite{Weinberg:1978kz,Gasser:1983yg},
offer a systematic framework to investigate the nature and composition of these challenging states.
They enable the inclusion of higher-order corrections in the interaction terms and make it possible to predict physical properties such as masses, decay widths, and scattering amplitudes.
Both the \XRes and \XResNovanta states have been investigated 
within perturbation theories based on chirally-motivated Lagrangians, starting from the first \XRes prediction in Ref.~\cite{ALK_Ramos}, followed by a subsequent work in Ref.~\cite{Garcia-Recio:ChiPtXi}, in which a second pole was found and assigned to the \XResNovanta.
EFT-based potentials involving elastic and inelastic transitions depend on the so-called low energy constants~\cite{Pich:EFT}, which must be determined from fits to the data. The number of these parameters increases as the potentials are extended to higher orders, hence larger data sets in wide energy ranges are needed as input, in particular for recent calculations at next-to-leading order (NLO)~\cite{Feijoo_KpXiK_1}.
In the strangeness sector, such experimental inputs are often scarce and challenging to obtain, due to the difficulties in performing experiments with unstable strange particles. 

Femtoscopy applied in high-energy \pp and nucleus-nucleus collisions has been proved to be a powerful experimental technique, able to provide precise data on hadron-hadron interactions~\cite{ALICE:pOmega,femtoreview}. 
Studies on the \Sminustwo meson-baryon interaction via measurements of \LAK correlations in \pp~\cite{ALICE:LAKpp} and \PbPb~\cite{LK_PbPbALICE} systems delivered for the first time evidence of the \XRes at the \LKMin channel threshold, offering a complementary approach to spectroscopy for investigating these molecular states by unveiling the interaction between their constituents~\cite{LambdaK_Vale}.
A further improvement in NLO Lagrangians, aimed at describing the \Sminustwo meson-baryon interaction and the nature of molecular candidates in this sector, requires precise experimental input on the \Xipi system~\cite{Feijoo:MagasXiPi,LambdaK_Vale2}, which is dynamically coupled to \LAK and provides crucial constraints on the coupled-channel dynamics.

The \Sminusone meson-baryon interaction is constrained by low-energy \KbarN scattering~\cite{Humphrey:1962zz, Watson:1963zz, Mast:1975pv, Nowak:1978au, Ciborowski:1982et,Piscicchia:2022wmd} and kaonic atom data~\cite{SIDDHARTA:2011dsy}, but theoretical models often exhibit discrepancies not only in describing the $\Lambda(1405)$ region below the \KbarN threshold but also at significantly higher energies.
ALICE used femtoscopy as an alternative method to study the \kbarN threshold interaction via \Kminusp~\cite{ALICE:pK,ALICE:pKCoupled,ALICE:pKPb}
correlations across \pp, \pPb and \PbPb collisions.
The femtoscopy measurements of \Kminusp pairs, in particular in small collision systems, demonstrated high sensitivity to the interplay among the different channels in this sector.
The region of energy well above the \kbarN threshold is more challenging to be explored experimentally since it involves systems with a neutral meson ($\upeta$) and a multi-strange baryon ($\Xi$), which are difficult to 
reconstruct due to low detection efficiency
or low production yield. Inelastic cross sections of \Kminusp $\rightarrow \upeta\Lambda$, $\upeta\Sigma^0$, $\mathrm{K}^0\Xi^0$ and $\mathrm{K}^+\Xi^-$ reactions are available, but are subject to large uncertainties, which impede the extraction of precise information on the underlying interaction~\cite{KpHigh1,KpHigh2,KpHigh3,KpHigh4,exp1,Mast:1975pv, exp3, Ciborowski:1982et, exp6, exp7, exp8, exp9, exp10, exp11}.
Dedicated studies based on ChPT calculations at NLO showed how incorporating data from high-energy channels leads to tighter constraints on the model
~\cite{Feijoo_KpXiK_1,Feijoo_KpXiK_2,Kamano:2014zba,Kamano:2015hxa}.
The \XiK channel is of particular relevance since it allows to investigate the isospin-triplet component of the \Sminusone meson-baryon interaction~\cite{Feijoo_KpXiK_3}, currently poorly constrained due to the limited availability of scattering data.
Precise data on the \XiK interaction~\cite{Doring:2025sgb}, in particular using femtoscopy~\cite{Feijoo_Femto_KpXiK}, are able to provide crucial experimental input for model discrimination across a wide energy range.

This work presents the first
measurement of femtoscopic correlations between $\Xi^-$ baryons (dss) and K$^{+}$ (u$\rm\overline{s}$)
and $\uppi^+$ 
(u$\rm\overline{d}$) mesons, and charge conjugates,
in high-multiplicity (HM) \pp collisions at \onethree, recorded by ALICE at the Large Hadron Collider.
The obtained correlations are compared with available models and the scattering parameters for the \XiK and \XiPi interaction are extracted through a fit using the \Ledn wavefunction including both the strong and Coulomb interactions.
In the fit to the \XiPi correlation function, the properties of \XRes and \XResNovanta states are determined.

\section{Data analysis}
\label{sec:DataAnalysis}

A total of $1\times10^9$ high-multiplicity \pp~collisions at $\sqrt{s}=13$~TeV were selected by ALICE~\cite{ALICE} using an online trigger based on the total signal amplitude measured by the V0 detector~\cite{VZERO}, which consists of two plastic scintillator arrays located on both sides of the collision point along the beam axis.
The selected HM events correspond to the 
0.17\%~highest-multiplicity events in the sample of inelastic collisions with at least one measured charged particle within the pseudorapidity range $|\eta|<1$ (INEL $>$ 0). 
The primary interaction vertex is reconstructed by extrapolating track segments from the innermost layers of the Inner Tracking System (ITS)~\cite{ALICEITS}, composed of six layers of silicon detectors providing high-precision position measurements, back to the beam line. Events with multiple reconstructed vertices are efficiently identified as in-bunch pile-up and rejected from the analysis~\cite{ALICEperf}; in the HM sample, the residual pile-up contamination is at the percent level and has a negligible impact on the final results.
These HM events have an average charged-particle multiplicity density at midrapidity $(|\eta|<0.5)$ of $\langle \mathrm{d}N_\mathrm{ch}/\mathrm{d}\eta\rangle \approx 30$, about 4 times larger than the INEL $>$ 0 events. 
It is also worth noting the enhanced production of hyperons in HM events~\cite{ALICE:EnhancedStrange} which facilitates the collection of large samples of the particle pairs of interest for this analysis. 

Charged particles are reconstructed and identified using the ITS, the Time Projection Chamber (TPC)~\cite{ALICETPC}, and the Time Of Flight detector (TOF)~\cite{ALICETOF}, which are immersed in a uniform magnetic field of 0.5 T along the beam direction and cover the pseudorapidity range $|\eta|<0.9$.
The TPC is a cylindrical gaseous detector that performs particle tracking and identification by measuring the specific energy loss (d$E$/d$x$).
The TOF detector consists of Multigap Resistive Plate Chambers and provides timing information.
Charged particles used for this analysis, including decay products from reconstructed $\Xi$ decays, are required to be within the pseudorapidity range $|\eta|<0.8$.
The transverse-momentum ($p_{\rm{T}}$) resolution for charged particles typically varies from about \SI{2}{\percent} for tracks with $\pt = 10$~GeV/$c$ to below \SI{1}{\percent} for $0.14<\pt<\SI{1}{}$~GeV/$c$.

Pions (kaons) are selected with a minimum transverse momentum of 0.14 (0.20)~\gevc.
Particle identification (PID) is performed using the d$E$/d$x$ measurement in the TPC and the timing information from the TOF, both as a function of the reconstructed track momentum.
The basic selection criteria for charged particles can be found in Ref.~\cite{ALICE:2023gxp}, and 
further details on the pion (kaon) selection are given in Refs.~\cite{ALICE:2024bhk,ALICE:CommonSourceppion} (\!\cite{PRX}). 
The purity of the pion and kaon sample was estimated to be 98\% and 100\%, respectively, using Monte Carlo (MC) simulations. The primary fraction, i.e. the fraction of particles stemming directly from the collision point, is calculated using the Monte Carlo template procedure described in Ref.~\cite{ALICE:Run1} and it amounted to $94\%$ for pions and $99\%$ for kaons.

The identification of charged $\Xi$ baryons exploits their characteristic weak decay cascade pattern. The primary decay is $\Xi^{-} \rightarrow \Lambda + \uppi^{-}$ 
, which has a branching ratio (BR) of nearly 100\%~\cite{PDG2024}. This is subsequently followed by the weak decay of the daughter $\Lambda$ baryon, $\Lambda \rightarrow$ p$ + \uppi^{-}$ (BR = 63.9 $\pm$ 0.5$\%$~\cite{PDG2024})
.
The selection of $\Xi$ candidates is performed in two main steps.
First, a pre-selection step applies loose, single-variable requirements on several topological quantities, as well as on the invariant masses of the $\Lambda$ and $\Xi$ candidates. This initial stage is designed to minimize signal loss rather than aggressively reduce combinatorial background.
In the second step, a multivariate analysis (MVA) is employed using a Boosted Decision Tree (BDT) technique, as detailed in Ref.~\cite{ALICE:2024rnr}.
A similar procedure was already used in previous femtoscopy analyses by ALICE involving D mesons~\cite{ALICE:2022enj,ALICE:2024bhk}.
A total of nine variables serve as input for the BDT. These include the distance of closest approach (DCA) between tracks, the DCA of tracks to the primary and decay vertices, the pointing angles of the $\Lambda$ and $\Xi$ candidates, and PID information for the charged particles.
The BDT is trained using MC simulations for the signal, where the production and decay of $\Xi$ baryons in HM pp events are simulated with PYTHIA. The reconstruction of $\Xi$ candidates is performed after propagating the daughter tracks through a full detector simulation incorporating the ALICE geometry and response using GEANT3~\cite{Brun:1994aa}.
The combinatorial background sample for the training is derived from data candidates that pass the pre-selection criteria but have a $\Xi$ invariant mass outside the signal region, specifically within the ranges 1280--1308 \mevcc and 1322--1370 \mevcc.
The application of the BDT selection significantly enhances the purity of the $\Xi$ sample while maintaining high signal efficiency. Following an optimization of the BDT output score threshold used to select signal candidates, the purity of the selected $\Xi$ sample reaches 97\%, with
minimal signal loss relative to the pre-selected data.
Finally, $\Xi$ candidates are retained if their invariant mass falls within a 10 \mevcc window centered around the nominal $\Xi^{-}$ mass.

The relative momentum between the selected particles in their pair rest frame, denoted as $k^*$, is measured for all combinations of interest. The experimental correlation function is then determined as 
\begin{equation}
    C(k^{*}) = \mathcal{N} \frac{N_{\text{same}}(k^{*})}{N_{\text{mixed}}(k^{*})},
    \label{eq:correlation_function} 
\end{equation}
where $N_{\text{same}}(k^{*})$ represents the $k^*$ distribution constructed from pairs of particles emitted within the same collision event. The $N_{\text{mixed}}(k^{*})$ distribution, which serves as a reference for uncorrelated pairs, is obtained by pairing particles from different events that exhibit similar primary vertex $z$-positions
and event multiplicities.
To account for the differing event sample sizes, the mixed-event distribution is normalized, via the constant $\mathcal{N}$, to the same-event distribution within the range \kstar $\in$ [1000--1200] \mevc, a region of the spectra where final state interaction (FSI) and resonance effects are absent.

The correlation functions are calculated for all oppositely charged particle combinations. 
No statistically significant differences are observed between the correlation functions of \XiMinK and \XiPK pairs, nor between \XiMinPi and \XiPPi pairs. 
Since the correlation functions are expected to be identical for charge-conjugate pairs, the respective pairs are combined to improve statistical precision. 
In the following, \XiK and \Xipi refer to the correlation functions obtained from summing the distributions of pairs and theirs charge-conjugates.

In the case of \Xipi correlations, autocorrelation effects are mitigated  by excluding pions identified as decay products of $\Xi$ baryons from the pairing.
Detector effects, such as track merging or track splitting, are evaluated by analyzing the $\Delta\eta - \Delta\varphi^{*}$ distributions, with $\varphi^{*}$ being the azimuthal coordinate between the $\Xi$ decay products and the paired meson at several radial positions within the TPC volume. These contributions are found to be negligible. 
Possible effects from combinatorial background due to out-of-bunch collision pile-up in the TPC are evaluated by repeating the analysis while requiring that all charged tracks have a matched hit in either the ITS 
or the TOF, which can provide precise timing information to remove pile-up contamination.
The effect is found to be negligible, and therefore, this requirement is not applied in the final analysis.

The measured \XiK and \Xipi correlation functions are presented in Figs.~\ref{fig:XiK} and ~\ref{fig:XiPiCFs}, respectively. The corresponding figures with extended \kstar range can be seen in the Appendix~\ref{app:appendix}.
The systematic uncertainties of the data points, depicted as shaded boxes in the figures, are estimated by varying the track selection criteria, following the procedure used in previous ALICE femtoscopy analyses, see Refs.~\cite{ALICE:Source_pionpion,PRX} for the procedure and for pion and kaon selections. For the $\Xi$ selection, the BDT threshold is varied within 25\%. The total systematic uncertainty remains below 1\% for both measured correlation functions in all \kstar bins.
Both the \XiK and \Xipi correlation functions show a clear signal of the FSI in the low relative momentum region, visible as a deviation from unity.
The \XiK correlation also shows 
a steadily increasing baseline, an enhancement that grows as \kstar decreases that is clearly visible in the region above 200 \MeVc, where FSI effects are negligible. This behavior contrasts with the \Xipi
data, which show no such trend and remain mostly constant in the large \kstar region.
Additionally, the \Xipi correlation shows a prominent peak at $\ks \approx 150$ \MeVc, corresponding to the invariant mass of $\approx$ 1530 \mevcc, where the signal of the \XResTrenta decaying into the \Xipi pairs with BR of $100\%$~\cite{PDG2024} is expected, and other structures. 
The features shown by both correlation functions are discussed in Secs.~\ref{sec:cf},~\ref{sec:ResultsXiK} and~\ref{sec:ResultsXipi}.

\section{Analysis of the correlation function}\label{sec:cf}

The measured correlation functions for \XiK and \XiPi pairs are fitted using the expression
\begin{align}\label{eq:totcorrelation}
    C_\mathrm{tot}(\kstar) = & N_{\mathrm{D}} \times C_\mathrm{\rm model}(\kstar) \times C_\mathrm{background} (\kstar),
\end{align}
where $N_{\mathrm{D}}$ is a normalization constant, left free to vary in the fit. Due to the different contribution of the background in the two pairs, discussed in details later in this section, the total fit range is $\ks \in [0,1000]$ \MeVc for \XiPi and $\ks \in [0,1600]$ \MeVc for \XiK.
The modeling term $C_{\rm model}(\ks)$ in Eq.~\ref{eq:totcorrelation} is analogous for both pairs and it is expressed as
\begin{align}\label{eq:Cmodel}
    C_{\rm model}(\ks) =
    \lambda_{\rm gen} C_{\rm gen}(\ks) 
    + \lambda_{\mathrm{M}-\Xi (1530) ^{\rm ch.}} C_{\mathrm{M}-\Xi (1530) ^{\rm ch.}}(\ks) 
        + \lambda_{\mathrm{M}-\tilde{\Xi}} C_{\mathrm{M}-\tilde{\Xi}}(\ks) 
    + \lambda_{\rm flat} C_{\rm flat}.
\end{align}
Each of the contributions entering in the modeled correlation is weighted by a corresponding $\lambda$ parameter, representing its relative fraction in the measured sample.

The first term $C_{\rm gen}(\ks)$ embeds the correlation signal stemming from the genuine FSI of meson-$\Xi$ pairs. The corresponding $\lambda_{\rm gen}$ amounts to 0.68 and 0.61 for \XiK and \Xipi pairs, respectively, and is determined as a product of the purity and fractions of the considered particles~\cite{ALICE:Run1}, provided in Sec.~\ref{sec:DataAnalysis}.

The second term, $C_{\mathrm{M}-\Xi(1530)^{\rm ch.}}(\ks)$, accounts for residual correlations arising when a primary pion/kaon (meson, M) is paired with a $\Xi$ originating from the decay of a charged (ch.) $\Xi(1530)$ resonance. 
A fraction of reconstructed $\Xi$ candidates indeed stems from the strong decay $\Xi(1530)^{-,0}\rightarrow \Xi^- \pi^{0,+}$, leading to a residual correlation that dilutes the genuine signal in the correlation function.
Since no theoretical predictions exist for the strong interaction between mesons and excited $\Xi(1530)$ states, only the contribution from the charged $\Xi(1530)^{-}$ ($\bar{\Xi}(1530)^+$) is modeled, assuming Coulomb interaction via the CATS framework~\cite{CATS}.
The relative \ks between the charged \XResTrenta and the meson is transformed, via the decay kinematics, 
into that of the measured \XiPi and \XiK pairs.
The corresponding weight $\lambda_{\mathrm{M}-\Xi (1530)^{\rm ch.}} = 0.14$ is evaluated following the same approach of Ref.~\cite{ALICE:LL} and is based on the measurement of the $\Xi(1530)^0$ yield in pp collisions at 7 TeV~\cite{ALICE:2014zxz}. 
In contrast, contributions from initial M–$\Xi(1530)^0$ pairs, with a relative contribution of 0.07, are assumed to be flat (equal to unity for any \kstar) and are absorbed into the $\lambda_{\rm flat}$ term (see below).

The third term in Eq.~\ref{eq:Cmodel}, $C_{\mathrm{M}-\tilde{\Xi}}(\ks)$, includes the contribution from primary mesons paired to misidentified $\Xi$ candidates and it amounts to $\lambda_{\mathrm{M}-\tilde{\Xi}}=0.03$. This residual contribution is modeled in a data-driven approach by calculating the correlation function using a selection of misidentified $\Xi$ candidates, realized by requiring a very low BDT score in the $\Xi$ selection.

Finally, the last term in Eq.~\ref{eq:Cmodel} refers to remaining contributions for which independence on \ks is assumed and hence $C_{\rm flat}$ is equal to unity, with a total weight $\lambda_{\rm flat}=0.15$.
The dominant contribution to $\lambda_{\rm flat}$ arises from \XResTrenta feed-down. The remaining contributions consist of pairs where pions or kaons are misidentified and, for \XiK pairs, kaons originating from the decay of $\phi$ mesons. The latter contribution is calculated consistently with previous femtoscopic studies involving kaons~\cite{ALICE:LAKpp,ALICE:2024bhk}.

To model the genuine $C_{\rm gen}(\ks)$ contribution, the \Ledn (LL) wavefunction approach for a charged pair is employed, including both strong and Coulomb interactions in s-wave~\cite{Lednicky:2005tb}.
The corresponding relative wave function reads
\begin{equation}
\label{eq:lednickyStrongCoulombWavefunction}
\psi(\vec{k}^*,\vec{r}^*) =e^{i\delta_c}\sqrt{A_{c}(\eta)}\left[e^{-i\vec{k}^*\cdot\vec{r}^*}F(-i\eta,1,i\xi)+f_{C}(k^*)\frac{\tilde{G}(\rho,\eta)}{r^*}
\right]\,.
\end{equation}
Here, $\eta=(k^*a_{C})^{-1}$ with $a_C$ being the Bohr radius and $\delta_c = \mathrm{Arg[\Gamma}(1+i\eta)]$ is the s-wave Coulomb phase-shift. The variable $\xi = \rho(1+\cos(\theta^*))$ is given in terms of $\rho = k^*r^*$, with $r^*$ representing the distance between the emitted particles and $\theta^*$ being the angle between $\vec k^*$ and $\vec r^*$. The asterisk refers to the evaluation in the pair rest frame. The term $A_C(\eta) = 2\pi\eta \left[ \exp(2\pi\eta) - 1 \right ]^{-1} $ represents the Coulomb penetration factor. The asymptotic expression of the Coulomb wave function is given by the term $e^{-\iota \vec k^*\vec \cdot r^*}F(\alpha,\, 1,\, z)$ together with $\widetilde{G}(\rho,\eta)$, where $F(\alpha,\, 1,\, z)$  is a confluent hypergeometric function and $\widetilde{G}(\rho,\eta) =  \sqrt{Ac}(G_0 + iF_0)$ is a combination of
the regular $(F_0)$ and singular $(G_0)$ $s$-wave Coulomb functions. The strong interaction is taken into account using the Coulomb-corrected scattering amplitude defined as 
\begin{equation}
\label{eq:amplitude_coulombstrong}
f_C(k^*)=\left[\frac{1}{f_0}+\frac{d_0{k^*}^{2}}{2}-ik^*A_C(k^*)-\frac{2}{a_C} h(k^*) \right]^{-1}\, ,
\end{equation}
with the effective range $d_0$ and the complex scattering length $f_0 = \Rescatt + i\cdot \Imscatt$. Here, the real and imaginary part of $f_0$ take into account the elastic and inelastic part on the interacting potential, respectively. The last term depends on the function $h(k^*)=\frac{1}{(k^*a_{C})^{2}}\sum^{\infty}_{n=1}\left[n\left(n^2+(k^*a_{C})^{-2}\right)\right]^{-1}-\gamma+\ln |k^*a_{C}|$  with the Euler constant $\gamma$ equal to $0.5772$.
In order to leave only the real (\Rescatt) and imaginary (\Imscatt) parts of the scattering length $f_0$ as free parameters to be determined from a fit to the data, the calculation is performed assuming a zero effective range $d_0$.
Note that there are known limitations of the LL model, particularly when dealing with the small source sizes characteristic of pp collisions and in presence of a rich coupled-channel dynamics, as in the systems under study~\cite{CATS,Albaladejo:2025kuv,AlbaladejoFemto,PRX,Torres-Rincon:2024znb,Feijoo:pRho}.
The LL approach has been employed in this study in order to provide an indication on the character and strength of strong interaction acting between $\Xi$ baryons and the light mesons.

In order to calculate the modeled correlations via the Koonin-Pratt formula~\cite{Lisa:2005dd,KOONIN197743}, information on the emitting source size and profile are needed. Following previous femtoscopic analyses performed in the same high-multiplicity data sample, the Resonance Source Model (RSM)~\cite{ALICE:Source,ALICE:Source_Corrigendum} is employed. The RSM is based on a data-driven approach in which the source is composed by a Gaussian core, common to all pairs and dependent only on the pair transverse mass \mt, and a non-Gaussian tail due to strongly-decaying resonances. The transverse mass of the pair is defined here as $\mt = \sqrt{k_{\mathrm{T}} ^2 + m^2}$, with pair transverse momentum $k_{\mathrm{T}}$ and average pair mass $m$. Recent works on same-charge $\uppi-\uppi$ and p$-\uppi$ pairs performed by the ALICE collaboration confirmed the validity of this approach~\cite{ALICE:Source_pionpion,ALICE:CommonSourceppion}. The core radius for \XiK pairs, determined following Ref.~\cite{ALICE:Source_Corrigendum}, is\rcoremTXiK\hspace{-0.2em}, and the one for \XiPi is\rcoremTXipi\hspace{-0.2em}. As observed already in previous femtoscopy studies~\cite{ALICE:2024bhk,ALICE:LAKpp}, 
the presence of long-lived strong resonances decaying into kaons and pions introduces a significant exponential tail to the source profile for large \rstar.
Consequently, a simple Gaussian function fails to describe the full source profile.
Following the approach used in the femtoscopy studies on $\Lambda \mathrm{K}^\pm$ in Ref.~\cite{ALICE:LAKpp}, the total core-halo source is therefore modeled with a weighted sum of two Gaussian functions, $S_1(\rstar)$ and $S_2(\rstar)$, weighted by the terms $\omega_\mathrm{S}$ and $\lambda_\mathrm{S}$, leading to an effective emitting source $S_\mathrm{eff}(\rstar)=\lambda_\mathrm{S}[\omega_\mathrm{S} S_1(\rstar)+(1-\omega_\mathrm{S})S_2(\rstar)]$. The values of the source parametrization for \XiK (\XiPi) pairs are \rUNOXiK, \rDOSXiK, \alphaSXiK, and \omegaSXiK~(\rUNOXipi, \rDOSXipi, \alphaSXipi, and \omegaSXipi).

A significant difference between the modeling of the \XiK, which occurs solely via Eq.~\ref{eq:Cmodel}, and the \XiPi pairs must be highlighted, and lies in the presence of several states in the latter. In particular, as can be seen in Fig.~\ref{fig:XiPiCFs}, the peak of the \XResTrenta is observed at $\ks \approx 145$ \MeVc. Additionally, the \XRes and \XResNovanta, expected to appear at $\ks\approx240$ \MeVc and at $\ks\approx303$ \MeVc, can be identified. Both these states lie close to the opening of the $\Lambda\mathrm{\bar{K}}^0$ ($\ks_{\Lambda\mathrm{\bar{K}}^0} = 232$ \MeVc) and $\Sigma^0\mathrm{\bar{K}}^0-\Sigma^\pm \mathrm{K}^\mp$ ($\ks_{\Sigma\mathrm{\bar{K}}} = 297-304$ \MeVc) channels coupled to the \XiPi system.
The current data do not permit an unambiguous identification between the channel openings and the resonant structures in this momentum region.
For the same reason, the $\Xi^0\upeta$ opening and the $\Xi(1950)$ state, at $\ks\approx 446$ \MeVc and $\ks\approx 513$ \MeVc respectively, cannot be discriminated from the combinatorial background. Since the two visible \XRes and \XResNovanta states carry as well information on the underlying \XiPi interaction, the modeled correlation for the \XiPi system includes as well these contributions, namely in the form
\begin{align}\label{eq:CmodelXiPi}
   C_{\mathrm{model}} ^{(\XiPi)}(\ks) =C_{\mathrm{model}}(\ks) + \sum_{i} w_i f_i (M_i, \Gamma_i),
\end{align}
where the sum $i$ runs over the three $\Xi^\ast$ states of mass $M_i$, width $\Gamma_i$, each described with a distribution $f_i$ weighted by the parameters $w_i$ to be determined in the fit. The narrow \XResTrenta is modeled assuming a Voigt distribution with mass and width given by the PDG values~\cite{PDG2024}.
The corresponding Gaussian broadening due to 
finite momentum resolution is also applied while performing the fit.
The broader \XRes and \XResNovanta states are modeled via a Breit-Wigner profile distribution with masses and widths extracted from the fit.

The final contribution needed to compare the total fit function in Eq.~\ref{eq:totcorrelation} with the measured correlations is the background $C_\mathrm{background} (\kstar)$. As already observed in several meson-baryon pairs~\cite{ALICE:pK,ALICE:pKCoupled,ALICE:pphi,ALICE:LAKpp,ALICE:2024bhk}, the dominant contribution to the non-femtoscopic signal is coming from the so-called mini-jet background. This background is typically associated with the initial hard processes occurring at the partonic level during the \pp collision~\cite{ALICE:minijet}, which can produce correlations visible up to \kstar values of around 1~\gevc or more. The effect of this background on the measured correlation has been observed to increase when the net-quantum numbers (baryon B, charge Q, strangeness S) within the pair are conserved~\cite{Vale_BBar}. Such a trend has been observed for example in the measured $\Lambda \mathrm{K}^+$ (S $ = 0$) and $\Lambda \mathrm{K}^-$ (\Sminustwo) correlations in Ref.~\cite{ALICE:LAKpp}. Therefore, a much larger contribution in the \XiK case,
in which an s$\rm\overline{s}$
pair is present in the final-state hadrons and can originate from a common partonic source,
with respect to the \XiPi one is expected. For both pairs, the $C_\mathrm{background} (\kstar)$ was built from Monte Carlo simulations based on the PYTHIA~8.2 generator~\cite{SJOSTRAND2015159} (Monash tune~\cite{Skands:2014pea}), and a simulation of the ALICE detector using GEANT3~\cite{Brun:1994aa}.
For the \XiPi system, the PYTHIA simulation provides a reasonable description of the data outside the femtoscopic region of $\ks \gtrsim 200$ \MeVc. The MC correlation is hence fitted in this case with a third-order degree polynomial in the range $\ks \in [0,1200]$ \MeVc and kept fixed in the fit of the experimental \Xipi correlation function.

As anticipated, the mini-jet background in the \XiK system has a larger influence on the shape of the correlation function.
The PYTHIA simulation captures the shape of the background correlation as a function of \kstar, but the strength of the correlation is underestimated as compared to data.
Hence, in order to  describe the data outside the femtoscopic region and extrapolate this baseline to the lower \kstar region, the correlation obtained from PYTHIA must be scaled.
For this, the \XiK background component is expressed as
\begin{align}\label{eq:CbackXiK}
C_\mathrm{background} ^{(\XiK)}(\kstar) = 1 + (C_{\mathrm{MC}}(\ks) - 1) \times S,
\end{align}
where $C_{\mathrm{MC}}(\ks)$ represents the PYTHIA correlation function and $S$ is a scaling factor parameter to be determined by fitting the data.
The PYTHIA correlation is modeled with a functional form including three Gaussian functions, with its parameters determined by a fit to the PYTHIA MC data in a large momentum range $\ks \in [0, 1600]$ \MeVc in order to constrain as much as possible the shape. 
The ALICE collaboration has reported substantial differences between PYTHIA tunes in the description of the angular correlations induced for \XiK pairs produced in pp collisions~\cite{ALICE:2023asw}.
In the current study, the Ropes tune~\cite{Bierlich:2022ned} was explored as an alternative to Monash; however, it did not lead to a substantial improvement in the overall description. The need to scale the correlation strength persists as a requirement to achieve a better fit to the data.

\section{Results on \texorpdfstring{\XiK}{XiK} correlations}\label{sec:ResultsXiK}

\begin{figure}[h!]
    \centering
    \includegraphics[width=0.89 \textwidth]{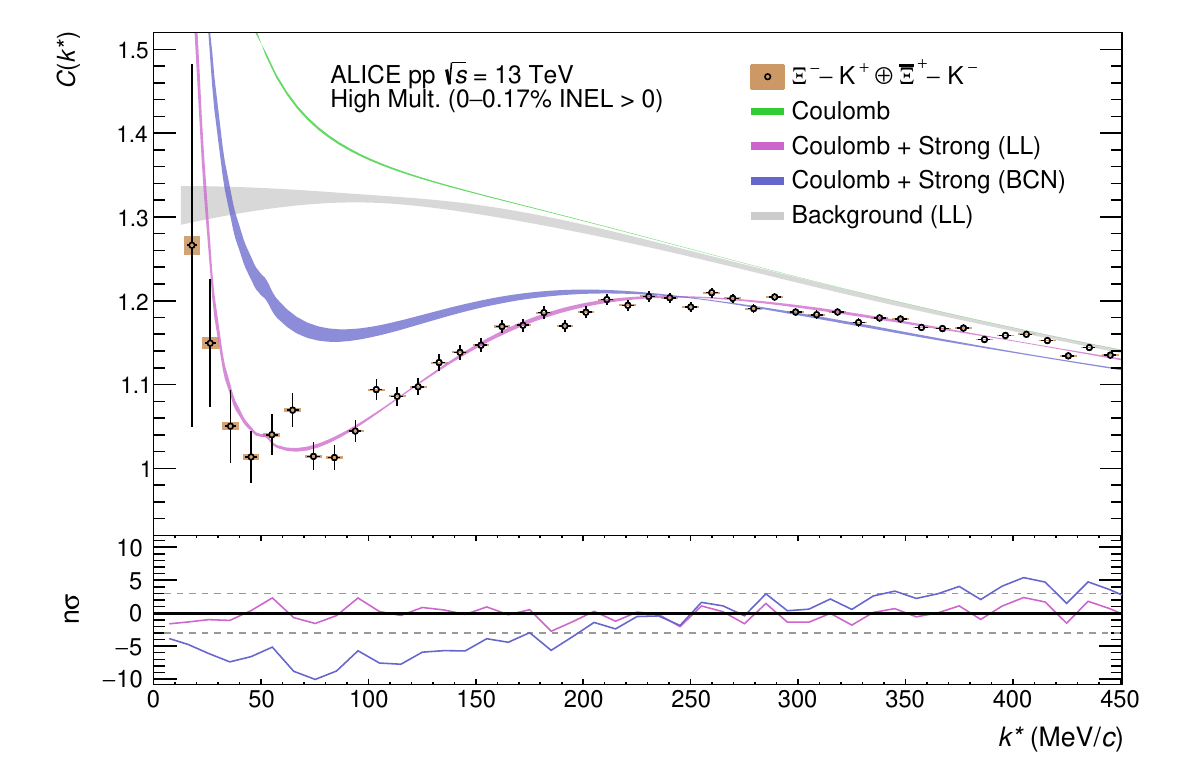}
    \caption{Measured correlation function of \XiK pairs. Data are represented by the black circles, with statistical (black vertical lines) and systematic (orange boxes) uncertainties shown separately.
    The pink and blue bands represent the results from the fit considering strong and Coulomb FSI via the LL approach and the BCN model, respectively.
    The gray band represents the mini-jet background associated with the LL fit.
    The green line represents the Coulomb-only assumption.
    The lower panel presents the difference between the data and the modeled correlation functions for the LL (pink) and BCN (blue) models expressed as the number of standard deviations, n$\sigma$, see text for details.
  (For interpretation of the references to colour in this figure legend, the reader is referred to the web version of this article.)
}
    \label{fig:XiK}
\end{figure}

The experimental correlation function for \XiK pairs is presented in Fig.~\ref{fig:XiK}. The correlation function is fitted to the model presented in Sec.~\ref{sec:cf} using the LL approach that incorporates the Coulomb interaction and a parametrization of the strong interaction based on the scattering parameters.
The fit is performed in the range
$\ks \in [0,1600]$ \MeVc.
The resulting model of the correlation function, according to Eq.~\ref{eq:Cmodel}, is shown as the pink band in Fig.~\ref{fig:XiK}.
The free parameters in this fit are \Rfzero and \Ifzero of the s-wave scattering length that enter in the LL model calculation of the wave-function, along with the parameter $S$ scaling the MC description of the mini-jet baseline contribution.
The mini-jet background alone is shown in Fig.~\ref{fig:XiK} by the gray band. 
The lower panel shows the difference between the data and the modeled correlation function as a function of \kstar, expressed in units of standard deviations ($n\sigma$), obtained by dividing the difference $C_\mathrm{data}-C_\mathrm{model}$ by the statistical uncertainty of the data.
See full fit range figure in the Appendix~\ref{app:appendix}.

The widths of the pink and gray bands reflect the systematic uncertainties propagated from the fit.
These uncertainties arise from several variations considered in the fitting procedure, namely:
i) variation of the upper limit of the fit range by $\pm 100$~\mevc;
ii) variation in the functional form used to model the mini-jet background as given by PYTHIA simulations, including just two Gaussian functions instead of three; 
iii) variations in the $\lambda$ parameters according to the uncertainties associated with their determination, which are at the level of 3\%;
iv) variations in the effective source radius according to the errors reported in the previous section. The fit is repeated for all possible combinations of these variations. At each \ks value, the width of the band corresponds to one standard deviation of the resulting distribution of solutions.

Fig.~\ref{fig:XiK} also displays the expectation from the Coulomb interaction alone, represented by the green line.
To build the Coulomb-only correlation function, intended just as a qualitative test, the data have not been fitted again, instead the effect of the genuine FSI from the LL fit has been substituted by a Coulomb-only calculation, hence the mini-jet background baseline determined from the full LL fit is used.
A clear difference is observed between the Coulomb-only expectation and the LL approach: the data lie significantly below the Coulomb-only curve, indicating that a repulsive strong interaction is necessary to describe the measured correlation function.
Indeed the scattering parameters obtained from the LL fit are 
\RfzeroXiK and \IfzeroXiK\hspace{-0.2em}.
Such rather large value of the real part of the scattering length reflects the necessity of a sizable repulsive interaction, while the value of the imaginary part shows a large influence of the inelastic channels, as expected for the \XiK system, being the last channel opening in the \Sminusone sector.

\begin{figure}[h!]
    \centering
    \includegraphics[width=0.89 \textwidth]{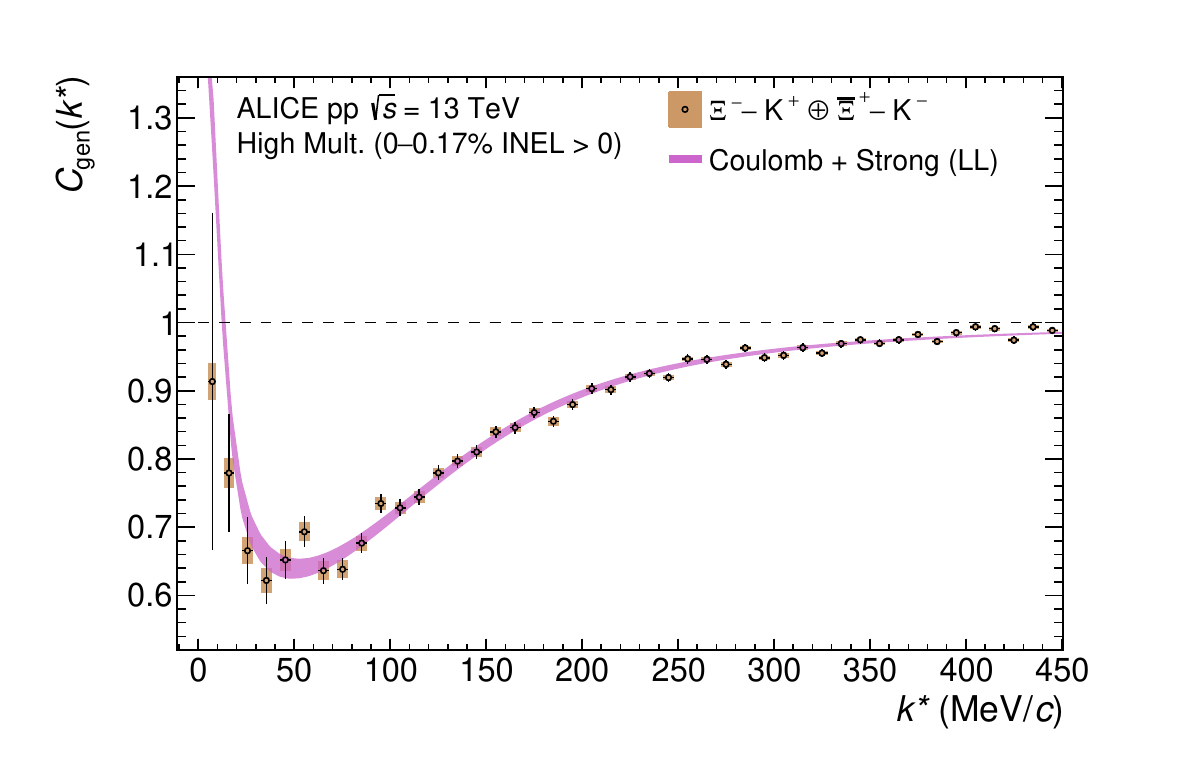}
    \caption{
    Extracted genuine \XiK correlation (data points) and genuine theoretical correlation function with the LL approach (pink line).
    Statistical (black vertical lines) and systematic (orange boxes) uncertainties of the data are shown separately.
    See text for details.
}
    \label{fig:XiK_genuine}
\end{figure}

Furthermore, Fig.~\ref{fig:XiK} includes a theoretical prediction from a  unitarized ChPT (UChPT) approach in coupled-channels, the BCN model~\cite{Feijoo_Femto_KpXiK}, shown as the blue line. This model derives unitarized s-wave meson-baryon scattering amplitudes from a chiral Lagrangian up to NLO. It is constrained to the available scattering data in the \Sminusone sector, including the higher-energy \Kminusp $\rightarrow$ \XiK reactions, and it is capable of describing the \Kminusp correlation function measured by ALICE in pp collisions~\cite{ALICE:pK,ALICE:pKCoupled} and the signature from the channel coupling in such data.
The BCN correlation function is calculated from these amplitudes using the Koonin-Pratt formula, applying the source function described in the previous section.
The mini-jet background parameter $S$ is also fitted when comparing the BCN model prediction to the data, although for clarity, the resulting baseline is not explicitly shown in the figure. 
The BCN model fails to describe the ALICE data in the low relative momentum region,
showing a weaker anticorrelation than observed in the data.
The scattering parameters for the \XiK interaction predicted by the BCN model, \Rfzero $= -0.39$ fm, \Ifzero $= 0.20$ fm, with an effective range of $d_{0} = 0.668 + i\cdot(-0.683)$ fm, are indeed smaller in absolute value compared with the ones obtained using the LL model under the zero effective range approximation.

Another coupled-channel calculation, under the Dynamical Coupled-Channels (DCC) approach by the ANL-Osaka group~\cite{Kamano:2014zba,Kamano:2015hxa}, which also starts from SU(3) Lagrangians to derive meson-baryon potentials, also delivers scattering parameters that differ from those obtained from the LL fit to the data. 
The \XiK isospin averaged scattering lengths from the ANL-Osaka approach are obtained considering two different models, with $\Rfzero = -0.76$ fm and \Ifzero = 0.11 fm , and  $\Rfzero = -0.86$ fm and \Ifzero = 0.09 fm.
In this case the real part of the scattering length is larger than the results from the LL fit while the imaginary part is smaller.
It is not surprising that the BCN and ANL-Osaka approaches deliver such different scattering parameters, even when both use the scattering data from reactions such as  \Kminusp $\rightarrow$ \XiK as input in the high energy regime.
The large uncertainties and limited availability of such data prevents from constraining the models at large energies~\cite{Doring:2025sgb} and this is reflected as well in the difficulties of describing the \XiK correlation data presented here.

Regarding a potential direct influence of resonant states on the \XiK correlation function, several theoretical frameworks predict indeed a plethora of $\Lambda^*$ and $\Sigma^*$ resonances that could couple to the \XiK channel across various energy ranges~\cite{hyperonI,hyperonII,hyperonIII,Zhang:2013sva,Kamano:2014zba,Kamano:2015hxa,Feijoo_KpXiK_2}.
However, the statistical precision of the current ALICE \XiK femtoscopic data is insufficient to discern clear, statistically significant structures in the correlation function that could be unambiguously attributed to the decay of such resonances into \XiK.
Consequently such resonant states are not considered in the fits to the experimental \XiK correlation function, although their undetected presence could influence the modeling of the measured correlations. 
Moreover, the possible presence of bound states or resonances below the \XiK threshold could induce an additional depletion of the correlation function. It is worth noting that the BCN model used for comparison with the data does not dynamically generate such states.

Figure~\ref{fig:XiK_genuine} shows a data-driven extracted genuine \XiK correlation function, obtained
by identifying $C_{\rm tot}(\ks)$
in Eq.\ref{eq:totcorrelation} with the experimental \XiK correlation function 
and isolating the $C_{\rm gen}(\ks)$ contribution in Eq.~\ref{eq:Cmodel}. 
The associated systematic uncertainties, depicted by the orange boxes, take into account both the data systematic uncertainties (see Sec.~\ref{sec:DataAnalysis}), and the systematic variations of the fit.
The genuine correlation function is compared with the genuine interaction part from the fit with the LL model (pink band) after consideration of the finite momentum resolution effects.

While the experimentally obtained \Cgen can be compared to any model of the \XiK interaction, given the source size reported in this work, it is important to note that its determination, and consequently the determination of the interaction parameters, inherently depends on the modeling of the mini-jet background baseline.
Since this baseline is determined simultaneously with the interaction parameters during the fit to the total correlation function, its characterization directly influences the extracted strong interaction signal represented by \Cgen.

\section{Results on \texorpdfstring{\XiPi}{XiPi} correlations}\label{sec:ResultsXipi}

\begin{figure}[h!]
    \centering
    \includegraphics[width=0.89 \textwidth]{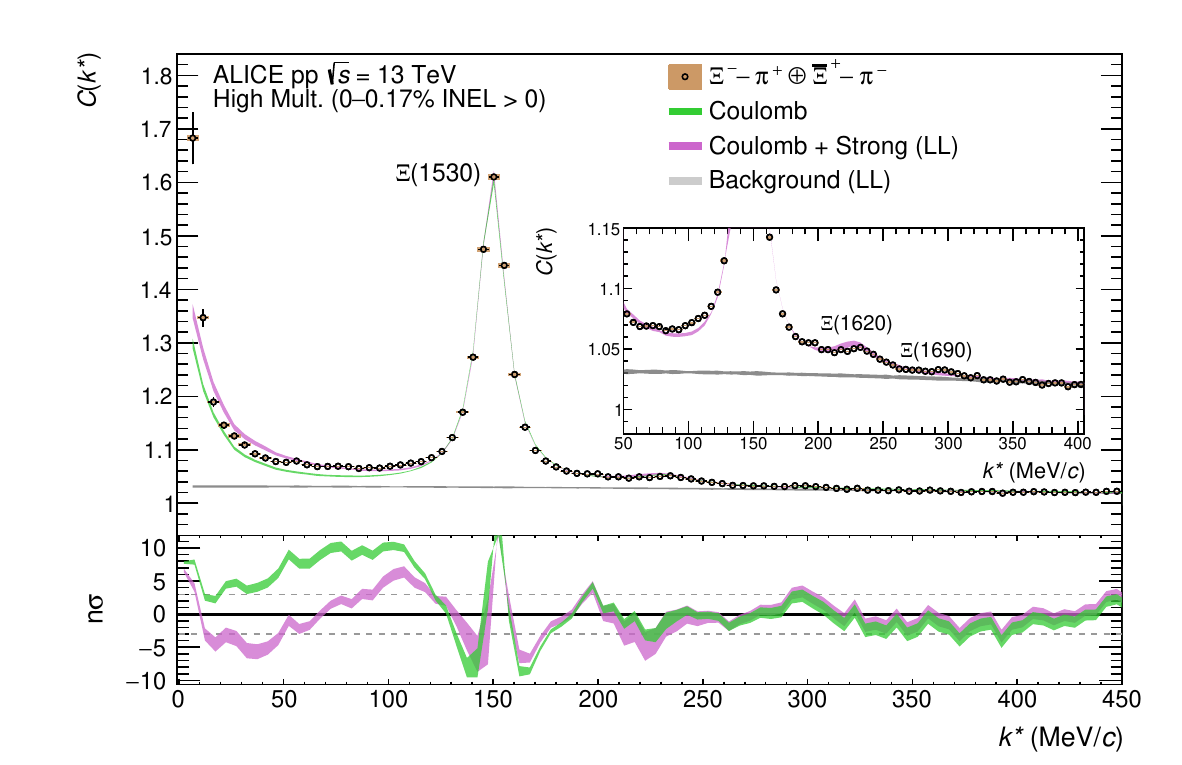}
    \caption{Measured correlation function of \XiPi pairs. Statistical (bars) and systematic (boxes) uncertainties are shown separately. The pink band represents the fit with both Coulomb and strong interactions via the LL approach, while the green band indicates the Coulomb-only results. The $C_\mathrm{background} (\kstar)$  within the Coulomb + strong fit is reported in gray. The inset shows a zoomed-in view of the y-axis, highlighting the \ks region of the \XRes and \XResNovanta states in greater detail.    
    The lower panel presents the difference between the data and the modeled correlation functions using Coulomb and strong interactions (pink) and Coulomb only (green) expressed as the number of standard deviations, n$\sigma$.  
    (For interpretation of the references to colour in this figure legend, the reader is referred to the web version of this article.)
    }
    \label{fig:XiPiCFs}
\end{figure}

The measured \XiPi correlation function is shown in Fig.~\ref{fig:XiPiCFs}. The pink and green bands depict the modeling of the total correlation function based on Eq.~\ref{eq:totcorrelation}, incorporating genuine, residuals and background contributions.
The influence of the $C_\mathrm{background} (\kstar)$ contribution, highlighted in gray and multiplied by $N_{\rm D}$, is clearly visible in the deviation of data from unity, starting already at large \kstar values.
The width of the bands represents the total uncertainty of the fit, estimated analogously to the \XiK case, except for the exclusion of variations on the background shape. Uncertainties on the \XiPi background are included by a bootstrap sampling from the MC correlation template~\cite{Bootstrap}. The fit range considered in the \XiPi case is $\ks \in [0,1000]$ \MeVc. See full fit range figure in the Appendix~\ref{app:appendix}.

The \XiPi correlation function lies above unity in the low \ks region, indicating an overall attractive interaction between $\Xi$ baryons and pions. A first scenario, shown by the green band, assumes a pure attraction coming from the Coulomb interaction and it clearly underestimates the data in the \ks region below the \XResTrenta peak. With the inclusion of the strong potential, via the \Ledn wavefunction approach discussed in Sec.~\ref{sec:cf}, a better description of the data is achieved, particularly in this low \ks region. The extracted value of $\Rescatt = 0.070 \pm 0.004(\mathrm{stat.})\pm 0.011(\mathrm{syst.})$ fm confirms the need for an additional shallow strong attraction to describe the data. The $\Imscatt = 0.002 \pm 0.003(\mathrm{stat.})\pm 0.010(\mathrm{syst.})$ is compatible with zero.
Such a result is expected since the \XiPi is the lowest energy channel that opens in the meson-baryon \Sminustwo sector. The values of $f_0$ obtained in this work are overall compatible with results on the \XiPi scattering length evaluated within UChPT in a coupled-channel approach in Ref.~\cite{Feijoo:MagasXiPi}. The improvement in the fit with the inclusion of a strong \XiPi interaction is estimated in terms of number of standard deviations $n\sigma$ between the data and the assumed model calculated over the full fit range. With the introduction of an additional strong attraction, the $n\sigma$ decreases to 15 from a value of 25 for the Coulomb-only scenario. A deviation between the assumed modeling and the data can be seen in the region below $\ks \approx 50$ \MeVc, with a clear overestimation of the measured correlation. Such a discrepancy might be caused by the need to include in the model higher-partial waves which could play a role in the \XiPi interacting potential already at low momenta, due to the appearance of the \XResTrenta state. Indeed, a deviation above the 5$\sigma$ level is also observed within the few data points around the \ks region of the \XResTrenta peak,  pointing as well to the need of an improved modeling of this state. More refined theoretical calculations, able to go beyond the s-wave assumption done in this work, could be useful in reducing the current disagreement.

As already discussed in Sec.~\ref{sec:cf}, two additional neutral states are visible in the measured \XiPi correlation, just above the \XResTrenta signal: the \XRes at $\ks\approx 240$ \MeVc and the \XResNovanta at $\ks\approx 300$ \MeVc. Both these states are shown in detail in the inset of Fig.~\ref{fig:XiPiCFs}. The properties of these two excited $\Xi^\ast$ have been determined from the fit that yields: $M_{\XRes} = 1608.16 \pm 1.56(\mathrm{stat.+syst.})$ \MeVmass, $\Gamma_{\XRes} = 47.12 \pm 5.63(\mathrm{stat.+syst.})$ MeV, $M_{\XResNovanta} = 1680.72 \pm 3.27(\mathrm{stat.+syst.})$ \MeVmass, $\Gamma_{\XResNovanta} = 26.41 \pm 13.86(\mathrm{stat.+syst.})$ MeV. The mass and width obtained for the neutral \XResNovanta state are in agreement with the reported PDG values~\cite{PDG2024} and with measurements by LHCb~\cite{LHCbXi16901820} and Belle~\cite{BelleXi} collaborations. For the much less known \XRes state, a lower mass value for the neutral state is extracted in comparison to the results obtained for the charged partner observed in the \LKMin correlation~\cite{ALICE:LAKpp}. Such a trend is expected, and widely observed, in isospin doublets based on the slight breaking of isospin symmetry~\cite{PDG2024}.
Compared with the Belle measurement in the same \XiPi channel, the mass value obtained from the \XiPi correlation is in agreement, while a narrower width is obtained, consistent with the results reported from \LKMin correlations~\cite{ALICE:LAKpp}.

Finally, it is worth mentioning a third structure present in the measured \XiPi correlation which appears at very large momenta. In the region of $850 < \ks < 950$ \MeVc, the peak of the produced $\Xi^{0}_{c}$ decaying into \XiPi (BR$=1.43\pm0.27\%$) is visible (see Fig.~\ref{fig:Xipi_fullrange} in Appendix~\ref{app:appendix}). In the same \ks region, the PDG reports another heavy excited $\Xi^\ast$ state, namely the \XResDuemila. 
Currently this state is very poorly known, with no precise information reported on mass, width and quantum numbers. A test of the feasibility of delivering quantitative constraints on the \XResDuemila has been performed; however, the current femtoscopy data can be described solely by the presence of the signal from the $\Xi^{0}_{c}$ decay. Additional details on this test are reported in the Appendix~\ref{app:appendix}.

\begin{figure}[h!]
    \centering
    \includegraphics[width=0.89 \textwidth]{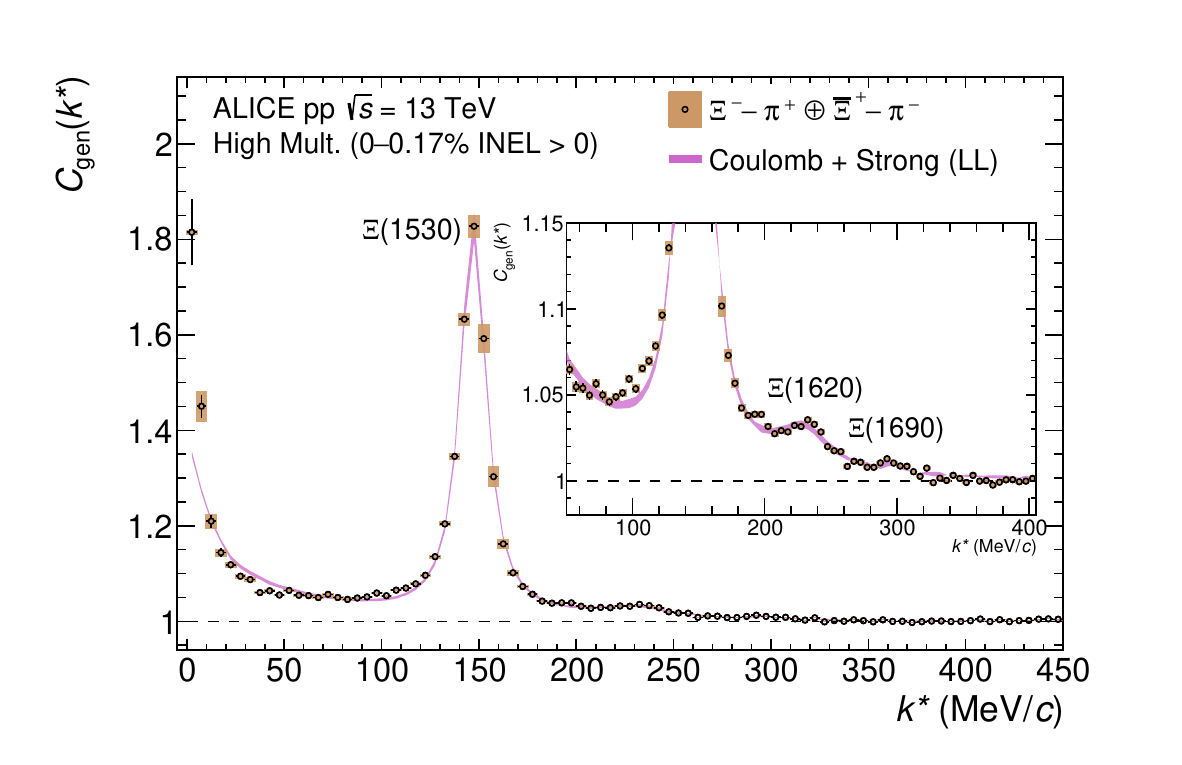}
    \caption{Genuine \XiPi correlation function. Statistical (bars) and systematic (boxes)
uncertainties are shown separately. The pink band represents the genuine theoretical correlation function obtained with both Coulomb and strong interactions. The inset shows a zoomed-in view of the y-axis, highlighting the \ks region of the \XRes and \XResNovanta states in greater detail.}
    \label{fig:XiPiCFs_genuine}
\end{figure}

Figure~\ref{fig:XiPiCFs_genuine} showcases the extracted genuine \XiPi correlation, obtained in an analogous way to the \XiK pairs.
The genuine \XiPi correlation is above unity, confirming the attractive nature of the \XiPi interaction, and it reaches unity just above the \XResNovanta state, indicating the robust treatment of the non-femtoscopic background thanks to the constrained MC template described in Sec.~\ref{sec:cf}. The reported pink band represents the genuine theoretical correlation, corrected for momentum resolution effects, obtained assuming the above-mentioned \XiPi scattering length. In the inset, the \XRes and \XResNovanta states are also shown and well reproduced by the considered model. The extraction of the genuine \XiPi correlation is free from the indetermination in the background description found in the case of \XiK, hence a model-independent set of high-precision data is delivered, covering a wide range of momenta, from the threshold up to 1 \gevc.

\section{Conclusions}
\label{sec:summary}
In this work, high-precision data on oppositely charged \XiK and \XiPi interactions are presented via measurements of their correlation functions in HM \pp collisions at \onethree. Scattering parameters for both systems are extracted using the \Ledn approximation for the wavefunction, indicating a repulsive strong interaction for \XiK and a shallow attraction for \XiPi. The data presented in this work constitute a fundamental input in understanding the underlying strong interaction between multi-strange and light hadrons.

The \XiK correlation function provides new
data in an energy region where direct scattering measurements are scarce.
These results therefore impose an important constraint for theoretical models aiming to describe the meson-baryon interaction in the \Sminusone sector.
So far, the observed level of repulsion in the \XiK system is not fully captured by the current theoretical approaches, highlighting the necessity of further refinement in the theoretical understanding of these interactions. 
It is important to highlight that the coupling to the $\Lambda\pi$ channel is predicted to have a substantial influence on the final \XiK correlation function~\cite{Feijoo_Femto_KpXiK}.
For this reason, it would be desirable to perform in the future a full coupled-channel analysis of the femtoscopy data in the \Sminusone sector, aiming to simultaneously describe the high-precision ALICE data for K$^{-}$p correlations together with the \XiK data here presented, and forthcoming $\Lambda\mbox{--}\pi$ correlations. 

The properties of \XRes and \XResNovanta states, visible in the measured \XiPi correlation function, are extracted and found to be overall compatible with previous measurements. Future measurements within the current LHC Run 3 and Run 4, aimed at a better understanding of the underlying coupled-channel dynamics in the \Sminustwo sector, should involve the challenging measurements of correlations involving $\Sigma\kbar$ pairs. Insights into these high energy channels could provide access to many heavy $\Xi^\ast$ states, still poorly known.

\newenvironment{acknowledgement}{\relax}{\relax}
\begin{acknowledgement}
\section*{Acknowledgements}
The ALICE Collaboration is grateful to Dr.~A. Feijoo and P. Encarnaci\'on for the fruitful discussions and valuable guidance on the theoretical description.

The ALICE Collaboration would like to thank all its engineers and technicians for their invaluable contributions to the construction of the experiment and the CERN accelerator teams for the outstanding performance of the LHC complex.
The ALICE Collaboration gratefully acknowledges the resources and support provided by all Grid centres and the Worldwide LHC Computing Grid (WLCG) collaboration.
The ALICE Collaboration acknowledges the following funding agencies for their support in building and running the ALICE detector:
A. I. Alikhanyan National Science Laboratory (Yerevan Physics Institute) Foundation (ANSL), State Committee of Science and World Federation of Scientists (WFS), Armenia;
Austrian Academy of Sciences, Austrian Science Fund (FWF): [M 2467-N36] and Nationalstiftung f\"{u}r Forschung, Technologie und Entwicklung, Austria;
Ministry of Communications and High Technologies, National Nuclear Research Center, Azerbaijan;
Rede Nacional de Física de Altas Energias (Renafae), Financiadora de Estudos e Projetos (Finep), Funda\c{c}\~{a}o de Amparo \`{a} Pesquisa do Estado de S\~{a}o Paulo (FAPESP) and The Sao Paulo Research Foundation  (FAPESP), Brazil;
Bulgarian Ministry of Education and Science, within the National Roadmap for Research Infrastructures 2020-2027 (object CERN), Bulgaria;
Ministry of Education of China (MOEC) , Ministry of Science \& Technology of China (MSTC) and National Natural Science Foundation of China (NSFC), China;
Ministry of Science and Education and Croatian Science Foundation, Croatia;
Centro de Aplicaciones Tecnol\'{o}gicas y Desarrollo Nuclear (CEADEN), Cubaenerg\'{\i}a, Cuba;
Ministry of Education, Youth and Sports of the Czech Republic, Czech Republic;
The Danish Council for Independent Research | Natural Sciences, the VILLUM FONDEN and Danish National Research Foundation (DNRF), Denmark;
Helsinki Institute of Physics (HIP), Finland;
Commissariat \`{a} l'Energie Atomique (CEA) and Institut National de Physique Nucl\'{e}aire et de Physique des Particules (IN2P3) and Centre National de la Recherche Scientifique (CNRS), France;
Bundesministerium f\"{u}r Bildung und Forschung (BMBF) and GSI Helmholtzzentrum f\"{u}r Schwerionenforschung GmbH, Germany;
General Secretariat for Research and Technology, Ministry of Education, Research and Religions, Greece;
National Research, Development and Innovation Office, Hungary;
Department of Atomic Energy Government of India (DAE), Department of Science and Technology, Government of India (DST), University Grants Commission, Government of India (UGC) and Council of Scientific and Industrial Research (CSIR), India;
National Research and Innovation Agency - BRIN, Indonesia;
Istituto Nazionale di Fisica Nucleare (INFN), Italy;
Japanese Ministry of Education, Culture, Sports, Science and Technology (MEXT) and Japan Society for the Promotion of Science (JSPS) KAKENHI, Japan;
Consejo Nacional de Ciencia (CONACYT) y Tecnolog\'{i}a, through Fondo de Cooperaci\'{o}n Internacional en Ciencia y Tecnolog\'{i}a (FONCICYT) and Direcci\'{o}n General de Asuntos del Personal Academico (DGAPA), Mexico;
Nederlandse Organisatie voor Wetenschappelijk Onderzoek (NWO), Netherlands;
The Research Council of Norway, Norway;
Pontificia Universidad Cat\'{o}lica del Per\'{u}, Peru;
Ministry of Science and Higher Education, National Science Centre and WUT ID-UB, Poland;
Korea Institute of Science and Technology Information and National Research Foundation of Korea (NRF), Republic of Korea;
Ministry of Education and Scientific Research, Institute of Atomic Physics, Ministry of Research and Innovation and Institute of Atomic Physics and Universitatea Nationala de Stiinta si Tehnologie Politehnica Bucuresti, Romania;
Ministerstvo skolstva, vyskumu, vyvoja a mladeze SR, Slovakia;
National Research Foundation of South Africa, South Africa;
Swedish Research Council (VR) and Knut \& Alice Wallenberg Foundation (KAW), Sweden;
European Organization for Nuclear Research, Switzerland;
Suranaree University of Technology (SUT), National Science and Technology Development Agency (NSTDA) and National Science, Research and Innovation Fund (NSRF via PMU-B B05F650021), Thailand;
Turkish Energy, Nuclear and Mineral Research Agency (TENMAK), Turkey;
National Academy of  Sciences of Ukraine, Ukraine;
Science and Technology Facilities Council (STFC), United Kingdom;
National Science Foundation of the United States of America (NSF) and United States Department of Energy, Office of Nuclear Physics (DOE NP), United States of America.
In addition, individual groups or members have received support from:
Czech Science Foundation (grant no. 23-07499S), Czech Republic;
FORTE project, reg.\ no.\ CZ.02.01.01/00/22\_008/0004632, Czech Republic, co-funded by the European Union, Czech Republic;
European Research Council (grant no. 950692), European Union;
Deutsche Forschungs Gemeinschaft (DFG, German Research Foundation) ``Neutrinos and Dark Matter in Astro- and Particle Physics'' (grant no. SFB 1258), Germany;
FAIR - Future Artificial Intelligence Research, funded by the NextGenerationEU program (Italy).

\end{acknowledgement}

\bibliographystyle{utphys}   
\bibliography{bibliography}

\newpage
\appendix

\section{Additional material}
\label{app:appendix} 

\subsection{Results on \texorpdfstring{$\Xi$(2500)}{Xi(2500)} state}

In this section, more details on the study performed on \XiPi pairs in the region of \ks between 830 and 920 \MeVc are provided, where a structure corresponding to the produced $\Xic \rightarrow \Xi\uppi$ resonance is visible. In the same region, an excited cascade, the \XResDuemila, is expected to be present. This state is currently poorly known, with one-star PDG rating~\cite{PDG2024}. The PDG is reporting only an approximate value of the mass ($\approx 2500$ \mevcc) and the related references are mainly experimental searches using kaon beams conducted in the early seventies and eighties~\cite{Aachen-Berlin-CERN-London-Vienna:1969bau,Xi25002,Xi25003}.

\begin{figure}[!htbp]
    \centering
    \includegraphics[width=\textwidth]{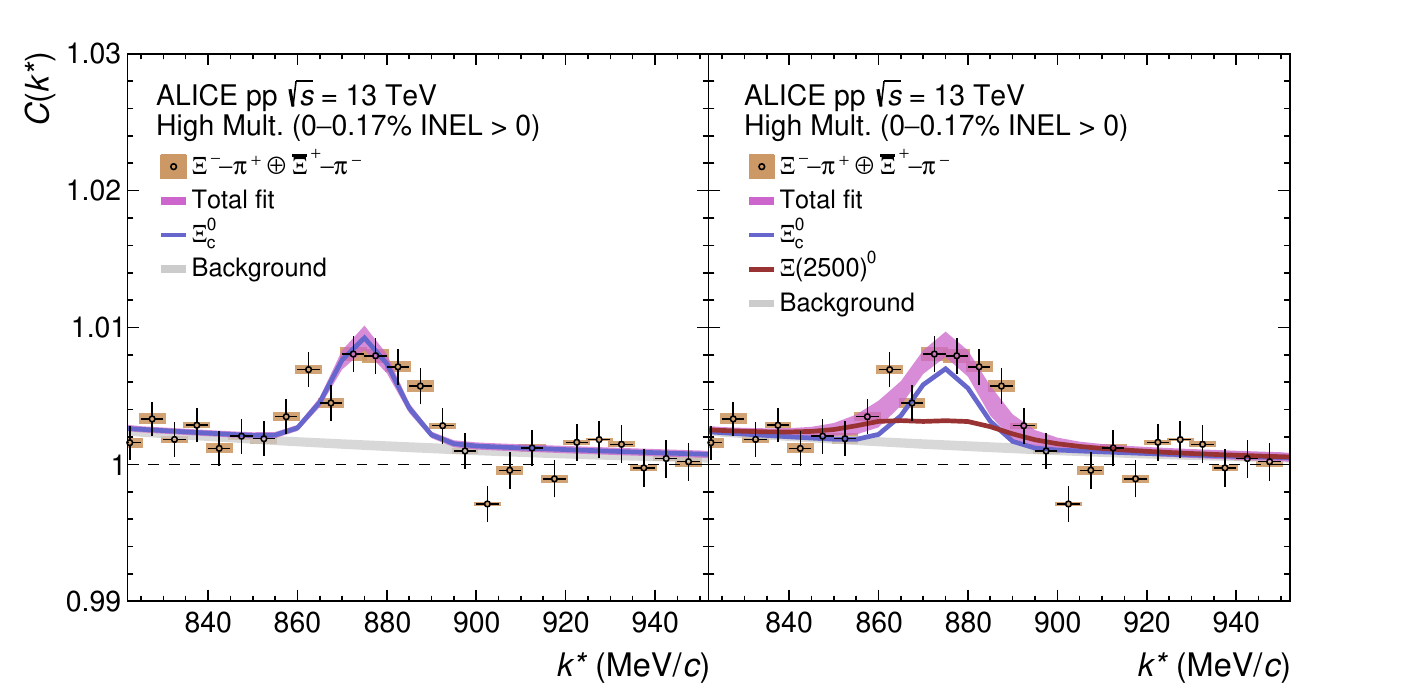}
    \caption{Region of the measured \XiPi correlation corresponding to the production of $\XicZero$ decaying into \XiPi pairs. Left: results assuming only the presence of the \Xic state. Right: results assuming also the presence of the \XResDuemila state. The pink band represents the fit with Eq.~\ref{eq:FitXic}, the polynomial background is given by the grey band. In blue and red, the center values of the profile shape of the \XResDuemila and \XicZero are shown, respectively.
    (For interpretation of the references to colour in this figure legend, the reader is referred to the web version of this article.)
    }
    \label{fig:CFXi2500}
\end{figure}

A chi-squared based likelihood test is conducted in the region $\ks \in[750-950]$\mevc by fitting the structure visible in Fig.~\ref{fig:CFXi2500} assuming either the only presence of the \Xic or adding as well the \XResDuemila state. In an attempt to extract properties of the \XResDuemila state, the known \Xic state is used as a constraint, i.e. it is included in the fit with fixed mass and width to its PDG nominal values~\cite{PDG2024}. Since the \Xic state is extremely narrow ($\Gamma \approx 10^{-6}$ MeV), a Monte Carlo simulation was performed in order to evaluate the effect of the \ks momentum resolution on the original distribution. The momentum smearing matrix, relating the original "true" $\ks_{\mathrm{true}}$ to the reconstructed $\ks_{\mathrm{reco}}$ one, is taken as initial input, limited to the region of interest in \ks. From the latter, at each value of $\ks_{\mathrm{true}}$, a corresponding Gaussian distribution in $\ks_{\mathrm{reco}}$ provides the mean $\langle\ks_{\mathrm{reco}}\rangle$ value and $\sigma_{\ks_{\mathrm{reco}}}$ dispersion. 
As expected, the width of the \XicZero is fully dominated by the resolution in \ks, found to be equal to $\Gamma_{\Xic} = \sigma_{\mathrm{res.}}=6.9$ \mevc.

The function used to perform the fit reads
\begin{align}\label{eq:FitXic}
   f_{\mathrm{fit}}(\ks) = B_{\mathrm{pol2}}(\ks) + A \cdot \left[ w_{\Xic} f_{\Xic} ^{G}(M_{\Xic},\Gamma_{\Xic}) + w_{\XResDuemila} f_{\XResDuemila} ^{BW}(M_{\XResDuemila},\Gamma_{\XResDuemila}) \right ], 
\end{align}

in which the free parameters to be determined in the fit are the ones related to the polynomial background $B_{\mathrm{pol2}}(\ks)=a_0(1+a_1\ks+a_2(\ks)^2)$, $A$, $w_{\Xic}$, $w_{\XResDuemila}$, $M_{\XResDuemila}$ and $\Gamma_{\XResDuemila}$. The \Xic is modeled via a Gaussian distribution with fixed width $\Gamma_{\Xic} = \sigma_{\mathrm{res.}}$ and fixed mass to the PDG value, $M_{\Xic} = 2470.4$ MeV (corresponding to $\ks = 874.7$ \MeVc)~\cite{PDG2024}. The corresponding distribution is shown in blue in both panels of Fig.~\ref{fig:CFXi2500}. A Breit-Wigner distribution is employed for the \XResDuemila, with mass $M_{\XResDuemila}$ and width $\Gamma_{\XResDuemila}$ to be extracted from the comparison to data.

Two scenarios have been tested in order to probe the sensitivity to the presence of the \XResDuemila state.
In the left panel of Fig.~\ref{fig:CFXi2500}, the fit assuming only the \Xic state is shown, using Eq.~\ref{eq:FitXic} with $w_{\XResDuemila} = 0$, and a reduced chi-square of $1.99$ is obtained. On the right panel of Fig.~\ref{fig:CFXi2500}, the results with the additional inclusion of the \XResDuemila state are presented for completeness. 
The limits of variation for the mass and width of the \XResDuemila are taken from the lower and upper limits provided by the previous measurements as listed by the PDG (see Refs.~\cite{Aachen-Berlin-CERN-London-Vienna:1969bau,Xi25002,Xi25003}).
In red, the obtained shape of the \XResDuemila state, scaled by its weight $w_{\XResDuemila}$, is also reported. The extracted  values of mass and width are $M_{\XResDuemila} = 2469.0\pm 14.3$ (stat. + syst.) \MeVmass and $\Gamma_{\XResDuemila}=32.0\pm22.2$ (stat. + syst.) MeV.
In this case, a reduced chi-square of $2.07$ is achieved, therefore indicating that within the current experimental precision, the presence of the \XResDuemila state is not necessary to describe the data. 

\subsection{Full fit range correlation functions for \texorpdfstring{\XiK}{XiK} and \texorpdfstring{\XiPi}{XiPi} pairs.}

 \begin{figure}[!htbp]
    \centering
    \includegraphics[width=0.89 \textwidth]{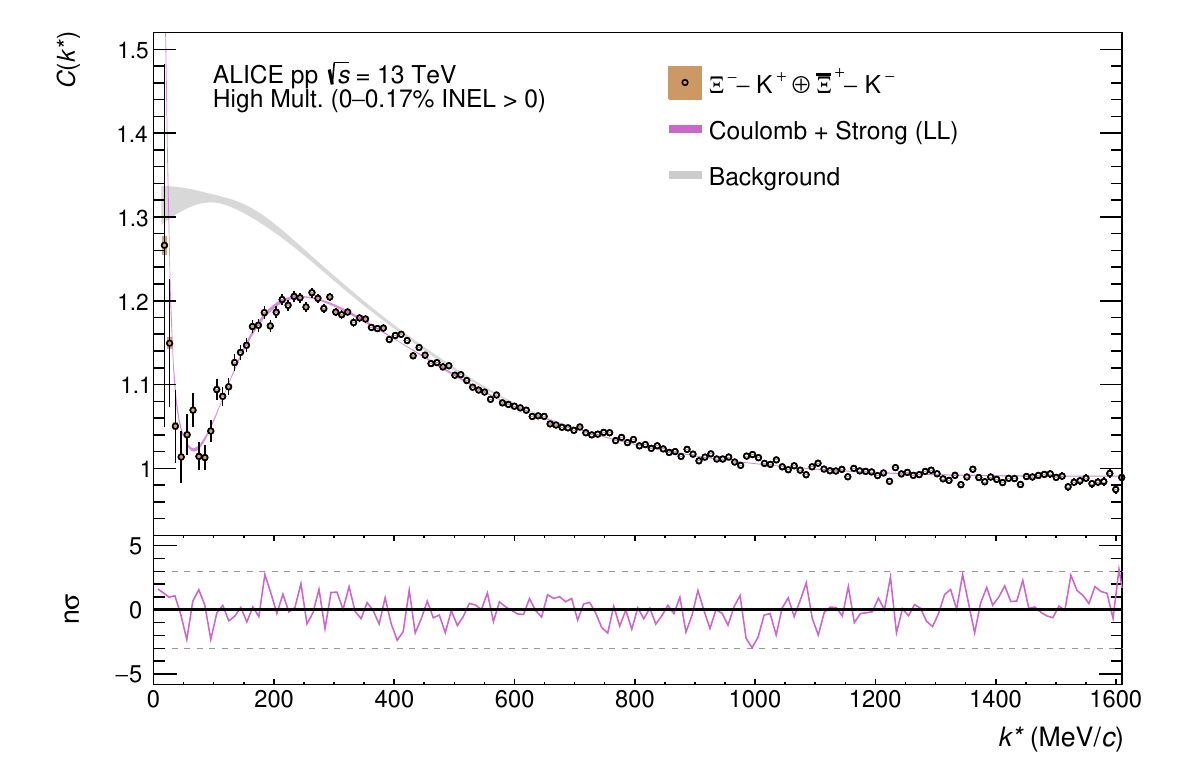}
    \caption{Measured correlation function of \XiK pairs in the whole fit range. Statistical (bars) and systematic (boxes)
uncertainties are shown separately. The pink band represents the fit with both Coulomb and strong interaction using the LL method. The $C_\mathrm{background} (\kstar)$ multiplied by the normalization constant $N_D$ obtained within the Coulomb + strong fit is reported in gray.}
    \label{fig:XiK_fullrange}
\end{figure}

 \begin{figure}[h!]
    \centering
    \includegraphics[width=0.89 \textwidth]{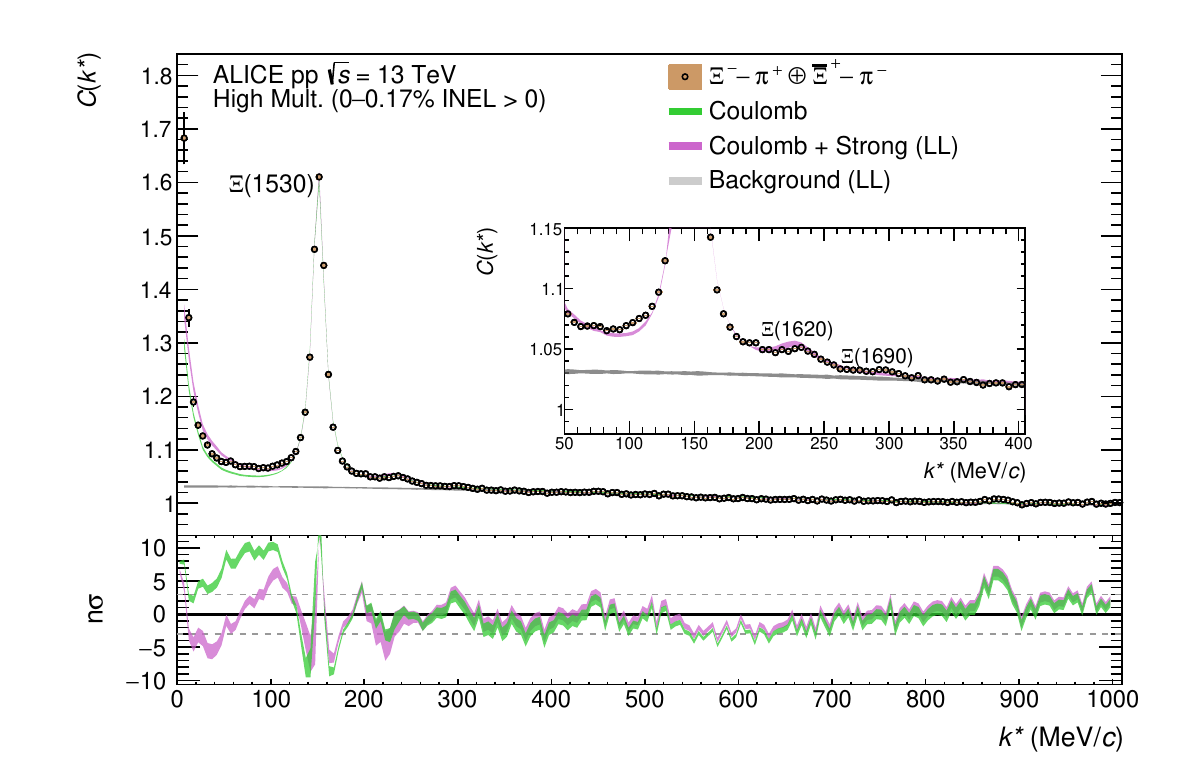
    }
    \caption{Measured correlation function of \XiPi pairs in the whole fit range. Statistical (bars) and systematic (boxes) uncertainties are shown separately. The pink band represents the fit with both Coulomb and strong interaction using the LL method. The $C_\mathrm{background} (\kstar)$ multiplied by the normalization constant $N_D$ obtained within the Coulomb + strong fit is reported in gray. The inset shows a zoomed-in view of the y-axis, highlighting the \ks region of the \XRes and \XResNovanta states in greater detail.}
    \label{fig:Xipi_fullrange}
\end{figure}

\clearpage
\section{The ALICE Collaboration}
\label{app:collab}
\begin{flushleft} 
\small

I.J.~Abualrob\,\orcidlink{0009-0005-3519-5631}\,$^{\rm 114}$, 
S.~Acharya\,\orcidlink{0000-0002-9213-5329}\,$^{\rm 50}$, 
G.~Aglieri Rinella\,\orcidlink{0000-0002-9611-3696}\,$^{\rm 32}$, 
L.~Aglietta\,\orcidlink{0009-0003-0763-6802}\,$^{\rm 24}$, 
N.~Agrawal\,\orcidlink{0000-0003-0348-9836}\,$^{\rm 25}$, 
Z.~Ahammed\,\orcidlink{0000-0001-5241-7412}\,$^{\rm 134}$, 
S.~Ahmad\,\orcidlink{0000-0003-0497-5705}\,$^{\rm 15}$, 
I.~Ahuja\,\orcidlink{0000-0002-4417-1392}\,$^{\rm 36}$, 
Z.~Akbar$^{\rm 81}$, 
A.~Akindinov\,\orcidlink{0000-0002-7388-3022}\,$^{\rm 140}$
% ...

\section*{Affiliation Notes}

$^{\rm I}$ Deceased\\
$^{\rm II}$ Also at: Max-Planck-Institut fur Physik, Munich, Germany\\
$^{\rm III}$ Also at: Czech Technical University in Prague (CZ)\\
$^{\rm IV}$ Also at: Instituto de Fisica da Universidade de Sao Paulo\\
$^{\rm V}$ Also at: Dipartimento DET del Politecnico di Torino, Turin, Italy\\
$^{\rm VI}$ Also at: Department of Applied Physics, Aligarh Muslim University, Aligarh, India\\
$^{\rm VII}$ Also at: Institute of Theoretical Physics, University of Wroclaw, Poland\\
$^{\rm VIII}$ Also at: Facultad de Ciencias, Universidad Nacional Aut\'{o}noma de M\'{e}xico, Mexico City, Mexico\\

\section*{Collaboration Institutes}

$^{1}$ A.I. Alikhanyan National Science Laboratory (Yerevan Physics Institute) Foundation, Yerevan, Armenia\\
$^{2}$ AGH University of Krakow, Cracow, Poland\\
$^{3}$ Bogolyubov Institute for Theoretical Physics, National Academy of Sciences of Ukraine, Kyiv, Ukraine\\
$^{4}$ Bose Institute, Department of Physics  and Centre for Astroparticle Physics and Space Science (CAPSS), Kolkata, India\\
$^{5}$ California Polytechnic State University, San Luis Obispo, California, United States\\
$^{6}$ Central China Normal University, Wuhan, China\\
$^{7}$ Centro de Aplicaciones Tecnol\'{o}gicas y Desarrollo Nuclear (CEADEN), Havana, Cuba\\
$^{8}$ Centro de Investigaci\'{o}n y de Estudios Avanzados (CINVESTAV), Mexico City and M\'{e}rida, Mexico\\
$^{9}$ Chicago State University, Chicago, Illinois, United States\\
$^{10}$ China Nuclear Data Center, China Institute of Atomic Energy, Beijing, China\\
$^{11}$ China University of Geosciences, Wuhan, China\\
$^{12}$ Chungbuk National University, Cheongju, Republic of Korea\\
$^{13}$ Comenius University Bratislava, Faculty of Mathematics, Physics and Informatics, Bratislava, Slovak Republic\\
$^{14}$ Creighton University, Omaha, Nebraska, United States\\
$^{15}$ Department of Physics, Aligarh Muslim University, Aligarh, India\\
$^{16}$ Department of Physics, Pusan National University, Pusan, Republic of Korea\\
$^{17}$ Department of Physics, Sejong University, Seoul, Republic of Korea\\
$^{18}$ Department of Physics, University of California, Berkeley, California, United States\\
$^{19}$ Department of Physics, University of Oslo, Oslo, Norway\\
$^{20}$ Department of Physics and Technology, University of Bergen, Bergen, Norway\\
$^{21}$ Dipartimento di Fisica, Universit\`{a} di Pavia, Pavia, Italy\\
$^{22}$ Dipartimento di Fisica dell'Universit\`{a} and Sezione INFN, Cagliari, Italy\\
$^{23}$ Dipartimento di Fisica dell'Universit\`{a} and Sezione INFN, Trieste, Italy\\
$^{24}$ Dipartimento di Fisica dell'Universit\`{a} and Sezione INFN, Turin, Italy\\
$^{25}$ Dipartimento di Fisica e Astronomia dell'Universit\`{a} and Sezione INFN, Bologna, Italy\\
$^{26}$ Dipartimento di Fisica e Astronomia dell'Universit\`{a} and Sezione INFN, Catania, Italy\\
$^{27}$ Dipartimento di Fisica e Astronomia dell'Universit\`{a} and Sezione INFN, Padova, Italy\\
$^{28}$ Dipartimento di Fisica `E.R.~Caianiello' dell'Universit\`{a} and Gruppo Collegato INFN, Salerno, Italy\\
$^{29}$ Dipartimento DISAT del Politecnico and Sezione INFN, Turin, Italy\\
$^{30}$ Dipartimento di Scienze MIFT, Universit\`{a} di Messina, Messina, Italy\\
$^{31}$ Dipartimento Interateneo di Fisica `M.~Merlin' and Sezione INFN, Bari, Italy\\
$^{32}$ European Organization for Nuclear Research (CERN), Geneva, Switzerland\\
$^{33}$ Faculty of Electrical Engineering, Mechanical Engineering and Naval Architecture, University of Split, Split, Croatia\\
$^{34}$ Faculty of Nuclear Sciences and Physical Engineering, Czech Technical University in Prague, Prague, Czech Republic\\
$^{35}$ Faculty of Physics, Sofia University, Sofia, Bulgaria\\
$^{36}$ Faculty of Science, P.J.~\v{S}af\'{a}rik University, Ko\v{s}ice, Slovak Republic\\
$^{37}$ Faculty of Technology, Environmental and Social Sciences, Bergen, Norway\\
$^{38}$ Frankfurt Institute for Advanced Studies, Johann Wolfgang Goethe-Universit\"{a}t Frankfurt, Frankfurt, Germany\\
$^{39}$ Fudan University, Shanghai, China\\
$^{40}$ Gangneung-Wonju National University, Gangneung, Republic of Korea\\
$^{41}$ Gauhati University, Department of Physics, Guwahati, India\\
$^{42}$ Helmholtz-Institut f\"{u}r Strahlen- und Kernphysik, Rheinische Friedrich-Wilhelms-Universit\"{a}t Bonn, Bonn, Germany\\
$^{43}$ Helsinki Institute of Physics (HIP), Helsinki, Finland\\
$^{44}$ High Energy Physics Group,  Universidad Aut\'{o}noma de Puebla, Puebla, Mexico\\
$^{45}$ Horia Hulubei National Institute of Physics and Nuclear Engineering, Bucharest, Romania\\
$^{46}$ HUN-REN Wigner Research Centre for Physics, Budapest, Hungary\\
$^{47}$ Indian Institute of Technology Bombay (IIT), Mumbai, India\\
$^{48}$ Indian Institute of Technology Indore, Indore, India\\
$^{49}$ INFN, Laboratori Nazionali di Frascati, Frascati, Italy\\
$^{50}$ INFN, Sezione di Bari, Bari, Italy\\
$^{51}$ INFN, Sezione di Bologna, Bologna, Italy\\
$^{52}$ INFN, Sezione di Cagliari, Cagliari, Italy\\
$^{53}$ INFN, Sezione di Catania, Catania, Italy\\
$^{54}$ INFN, Sezione di Padova, Padova, Italy\\
$^{55}$ INFN, Sezione di Pavia, Pavia, Italy\\
$^{56}$ INFN, Sezione di Torino, Turin, Italy\\
$^{57}$ INFN, Sezione di Trieste, Trieste, Italy\\
$^{58}$ Inha University, Incheon, Republic of Korea\\
$^{59}$ Institute for Gravitational and Subatomic Physics (GRASP), Utrecht University/Nikhef, Utrecht, Netherlands\\
$^{60}$ Institute of Experimental Physics, Slovak Academy of Sciences, Ko\v{s}ice, Slovak Republic\\
$^{61}$ Institute of Physics, Homi Bhabha National Institute, Bhubaneswar, India\\
$^{62}$ Institute of Physics of the Czech Academy of Sciences, Prague, Czech Republic\\
$^{63}$ Institute of Space Science (ISS), Bucharest, Romania\\
$^{64}$ Institut f\"{u}r Kernphysik, Johann Wolfgang Goethe-Universit\"{a}t Frankfurt, Frankfurt, Germany\\
$^{65}$ Instituto de Ciencias Nucleares, Universidad Nacional Aut\'{o}noma de M\'{e}xico, Mexico City, Mexico\\
$^{66}$ Instituto de F\'{i}sica, Universidade Federal do Rio Grande do Sul (UFRGS), Porto Alegre, Brazil\\
$^{67}$ Instituto de F\'{\i}sica, Universidad Nacional Aut\'{o}noma de M\'{e}xico, Mexico City, Mexico\\
$^{68}$ iThemba LABS, National Research Foundation, Somerset West, South Africa\\
$^{69}$ Jeonbuk National University, Jeonju, Republic of Korea\\
$^{70}$ Korea Institute of Science and Technology Information, Daejeon, Republic of Korea\\
$^{71}$ Laboratoire de Physique Subatomique et de Cosmologie, Universit\'{e} Grenoble-Alpes, CNRS-IN2P3, Grenoble, France\\
$^{72}$ Lawrence Berkeley National Laboratory, Berkeley, California, United States\\
$^{73}$ Lund University Department of Physics, Division of Particle Physics, Lund, Sweden\\
$^{74}$ Marietta Blau Institute, Vienna, Austria\\
$^{75}$ Nagasaki Institute of Applied Science, Nagasaki, Japan\\
$^{76}$ Nara Women{'}s University (NWU), Nara, Japan\\
$^{77}$ National and Kapodistrian University of Athens, School of Science, Department of Physics , Athens, Greece\\
$^{78}$ National Centre for Nuclear Research, Warsaw, Poland\\
$^{79}$ National Institute of Science Education and Research, Homi Bhabha National Institute, Jatni, India\\
$^{80}$ National Nuclear Research Center, Baku, Azerbaijan\\
$^{81}$ National Research and Innovation Agency - BRIN, Jakarta, Indonesia\\
$^{82}$ Niels Bohr Institute, University of Copenhagen, Copenhagen, Denmark\\
$^{83}$ Nikhef, National institute for subatomic physics, Amsterdam, Netherlands\\
$^{84}$ Nuclear Physics Group, STFC Daresbury Laboratory, Daresbury, United Kingdom\\
$^{85}$ Nuclear Physics Institute of the Czech Academy of Sciences, Husinec-\v{R}e\v{z}, Czech Republic\\
$^{86}$ Oak Ridge National Laboratory, Oak Ridge, Tennessee, United States\\
$^{87}$ Ohio State University, Columbus, Ohio, United States\\
$^{88}$ Physics department, Faculty of science, University of Zagreb, Zagreb, Croatia\\
$^{89}$ Physics Department, Panjab University, Chandigarh, India\\
$^{90}$ Physics Department, University of Jammu, Jammu, India\\
$^{91}$ Physics Program and International Institute for Sustainability with Knotted Chiral Meta Matter (WPI-SKCM$^{2}$), Hiroshima University, Hiroshima, Japan\\
$^{92}$ Physikalisches Institut, Eberhard-Karls-Universit\"{a}t T\"{u}bingen, T\"{u}bingen, Germany\\
$^{93}$ Physikalisches Institut, Ruprecht-Karls-Universit\"{a}t Heidelberg, Heidelberg, Germany\\
$^{94}$ Physik Department, Technische Universit\"{a}t M\"{u}nchen, Munich, Germany\\
$^{95}$ Politecnico di Bari and Sezione INFN, Bari, Italy\\
$^{96}$ Research Division and ExtreMe Matter Institute EMMI, GSI Helmholtzzentrum f\"ur Schwerionenforschung GmbH, Darmstadt, Germany\\
$^{97}$ Saga University, Saga, Japan\\
$^{98}$ Saha Institute of Nuclear Physics, Homi Bhabha National Institute, Kolkata, India\\
$^{99}$ School of Physics and Astronomy, University of Birmingham, Birmingham, United Kingdom\\
$^{100}$ Secci\'{o}n F\'{\i}sica, Departamento de Ciencias, Pontificia Universidad Cat\'{o}lica del Per\'{u}, Lima, Peru\\
$^{101}$ SUBATECH, IMT Atlantique, Nantes Universit\'{e}, CNRS-IN2P3, Nantes, France\\
$^{102}$ Sungkyunkwan University, Suwon City, Republic of Korea\\
$^{103}$ Suranaree University of Technology, Nakhon Ratchasima, Thailand\\
$^{104}$ Technical University of Ko\v{s}ice, Ko\v{s}ice, Slovak Republic\\
$^{105}$ The Henryk Niewodniczanski Institute of Nuclear Physics, Polish Academy of Sciences, Cracow, Poland\\
$^{106}$ The University of Texas at Austin, Austin, Texas, United States\\
$^{107}$ Universidad Aut\'{o}noma de Sinaloa, Culiac\'{a}n, Mexico\\
$^{108}$ Universidade de S\~{a}o Paulo (USP), S\~{a}o Paulo, Brazil\\
$^{109}$ Universidade Estadual de Campinas (UNICAMP), Campinas, Brazil\\
$^{110}$ Universidade Federal do ABC, Santo Andre, Brazil\\
$^{111}$ Universitatea Nationala de Stiinta si Tehnologie Politehnica Bucuresti, Bucharest, Romania\\
$^{112}$ University of Cape Town, Cape Town, South Africa\\
$^{113}$ University of Derby, Derby, United Kingdom\\
$^{114}$ University of Houston, Houston, Texas, United States\\
$^{115}$ University of Jyv\"{a}skyl\"{a}, Jyv\"{a}skyl\"{a}, Finland\\
$^{116}$ University of Kansas, Lawrence, Kansas, United States\\
$^{117}$ University of Liverpool, Liverpool, United Kingdom\\
$^{118}$ University of Science and Technology of China, Hefei, China\\
$^{119}$ University of Silesia in Katowice, Katowice, Poland\\
$^{120}$ University of South-Eastern Norway, Kongsberg, Norway\\
$^{121}$ University of Tennessee, Knoxville, Tennessee, United States\\
$^{122}$ University of the Witwatersrand, Johannesburg, South Africa\\
$^{123}$ University of Tokyo, Tokyo, Japan\\
$^{124}$ University of Tsukuba, Tsukuba, Japan\\
$^{125}$ Universit\"{a}t M\"{u}nster, Institut f\"{u}r Kernphysik, M\"{u}nster, Germany\\
$^{126}$ Universit\'{e} Clermont Auvergne, CNRS/IN2P3, LPC, Clermont-Ferrand, France\\
$^{127}$ Universit\'{e} de Lyon, CNRS/IN2P3, Institut de Physique des 2 Infinis de Lyon, Lyon, France\\
$^{128}$ Universit\'{e} de Strasbourg, CNRS, IPHC UMR 7178, F-67000 Strasbourg, France, Strasbourg, France\\
$^{129}$ Universit\'{e} Paris-Saclay, Centre d'Etudes de Saclay (CEA), IRFU, D\'{e}partment de Physique Nucl\'{e}aire (DPhN), Saclay, France\\
$^{130}$ Universit\'{e}  Paris-Saclay, CNRS/IN2P3, IJCLab, Orsay, France\\
$^{131}$ Universit\`{a} degli Studi di Foggia, Foggia, Italy\\
$^{132}$ Universit\`{a} del Piemonte Orientale, Vercelli, Italy\\
$^{133}$ Universit\`{a} di Brescia, Brescia, Italy\\
$^{134}$ Variable Energy Cyclotron Centre, Homi Bhabha National Institute, Kolkata, India\\
$^{135}$ Warsaw University of Technology, Warsaw, Poland\\
$^{136}$ Wayne State University, Detroit, Michigan, United States\\
$^{137}$ Yale University, New Haven, Connecticut, United States\\
$^{138}$ Yildiz Technical University, Istanbul, Turkey\\
$^{139}$ Yonsei University, Seoul, Republic of Korea\\
$^{140}$ Affiliated with an institute formerly covered by a cooperation agreement with CERN\\
$^{141}$ Affiliated with an international laboratory covered by a cooperation agreement with CERN.\\

\end{flushleft}

\end{document}